\documentclass{aa}  

\usepackage{graphicx}
\usepackage{txfonts}
\usepackage{siunitx}						
\usepackage{physics}
\usepackage{hyperref}
\usepackage{multicol}
\usepackage{lipsum}
\usepackage{ulem}
\usepackage{graphicx}
\usepackage{subcaption}
\usepackage{mwe}
\usepackage{tikz}
\usetikzlibrary{positioning}
\usepackage{balance}

\usepackage{xcolor}
\newcommand{\arwv}{\texttt{ARWV }}
\newcommand{\reb}{\texttt{REBOUND }}
\newcommand{\rebx}{\texttt{REBOUNDX }}

\definecolor{burntorange}{rgb}{0.8, 0.33, 0.0}

\hypersetup{
	colorlinks			= true			, 
	citecolor			= blue!80!black			, 
	linkcolor			= blue			, 
	urlcolor			=blue!80!black		, 
	breaklinks			=true           ,
}

\usepackage{natbib}
\bibpunct{(}{)}{;}{a}{}{,}             
\begin{document}

\title{Impact of a granular mass distribution on the orbit of S2 in the Galactic center}

\author{
    M.~Sadun Bordoni \inst{1}\thanks{mbordoni{@}mpe.mpg.de} \and
    R.~Capuzzo Dolcetta \inst{2,8,9,10} \and
    A.~Generozov \inst{3,4} \and
    G.~Bourdarot \inst{1} \and
    A.~Drescher \inst{1} \and
    F.~Eisenhauer \inst{1,5} \and
    R.~Genzel \inst{1,6}
    S.~Gillessen \inst{1} \and
    S.~Joharle \inst{1} \and
    F.~Mang \inst{1} \and
    T.~Ott \inst{1} \and
    D.C.~Ribeiro \inst{1} \and
    S.D.~von Fellenberg \inst{1,7} 
}
  
\institute{
  	    Max Planck Institute for Extraterrestrial Physics, Giessenbachstraße 1, 85748 Garching, Germany \and	
            Department of Physics, Sapienza, Universit\'a di Roma, P.le A. Moro 5, 00185 Rome, Italy \and	
            Technion - Israel Institute of Technology Haifa, 3200003, Israel \and	
            Astronomy Dept and Oden Institute, University of Texas at Austin, Austin, TX 78712, USA\and
  	    Department of Physics, Technical University of Munich, 85748 Garching, Germany \and
  	    Departments of Physics \& Astronomy, Le Conte Hall, University of California, Berkeley, CA 94720, USA \and
            Max-Planck-Institute für Radioastronomie, Auf dem Hügel 69, D-53121 Bonn, Germany \and
            Centro Ricerche Enrico Fermi, Via Panisperna 89a, 00184 Rome, Italy \and
            Istituto Nazionale Fisica Nucleare, Unit\'a Roma 1, Dipartimento di Fisica, Sapienza, Universit\'a di Roma, P.le A.Moro 5, 00185 Rome, Italy \and
            Istituto Nazionale di Astrofisica, Osservatorio astronomico di Roma (Monteporzio, Italy) 
} 

\date{Accepted for publication in Astronomy \& Astrophysics on 2025-06-28}

\abstract{
The orbit of the S2 star around Sagittarius A* provides a unique opportunity to test general relativity and study dynamical processes near a supermassive black hole. Observations have shown that the orbit of S2 is consistent with a Schwarzschild orbit at a 10$\sigma$ confidence level, constraining the amount of extended mass within its orbit to less than 1200 M$_\odot$, under the assumption of a smooth, spherically symmetric mass distribution.
In this work we investigate the effects on the S2 orbit of \textit{granularity} in the mass distribution, assuming it consists of a cluster of equal-mass objects surrounding Sagittarius A*.
Using a fast dynamical approach validated by full $N$-body simulations, we perform a large set of simulations of the motion of S2 with different realizations of the cluster objects distribution. We find that granularity can induce significant deviations from the orbit in case of a smooth potential, causing precession of the orbital plane and a variation of the in-plane precession.
Larger deviations are observed for higher masses of individual cluster objects and increased total mass of the cluster.
For a cluster mass of 1000 M$_\odot$ enclosed within the apocenter of S2, the average precession of the S2 orbital plane over a full orbit reaches up to 0.2 arcmin for 1 M$_\odot$ cluster objects and up to 1.5 arcmin for 100 M$_\odot$ objects. The in-plane precession deviates by up to 1.5 arcmin, corresponding to a fractional variation of 13\%.
Interactions with the cluster objects also induce a sort of `Brownian motion' of Sagittarius A*, with a mean displacement of up to 6 \text{$\mu as$} and velocity up to 238 m s$^{-1}$.
Mock data analysis reveals that these effects could produce observable deviations in the trajectory of S2 from a Schwarzschild orbit, especially near apocenter. During the next apocenter passage of S2 in 2026, astrometric residuals in Declination may exceed the astrometric accuracy threshold of GRAVITY ($\approx$ 30 \text{$\mu as$}), as it happens in 35 to 60\% of simulations for black holes of 20 to 100 M$_\odot$. This presents a unique opportunity to detect, for the first time, scattering effects on the orbit of S2 caused by stellar-mass black holes, thanks to the remarkable precision achievable with GRAVITY and its future upgraded version, GRAVITY+.
We also demonstrate that any attempt to constrain the extended mass enclosed within the orbit of S2 must explicitly account for granularity in the stellar‐mass black hole population.}
  

\titlerunning{Impact of a granular mass distribution on the orbit of S2 in the Galactic center}
\authorrunning{Sadun Bordoni et al.}
\maketitle


 \section{Introduction}
\label{introduction}

The S-stars orbiting the supermassive black hole (SMBH) Sagittarius A* (Sgr A*) in the Galactic center (GC) offer a unique laboratory for studying the dynamics of stars in an intermediate-strong gravitational field. Among these stars, S2 has played a central role due to its relatively short orbital period of about 16 years and its brightness ($m_K \approx 14$). Over the past three decades, its motion around Sgr A* has been monitored through adaptive optics assisted astrometry and spectroscopy at the Very Large Telescope (VLT) and the Keck Observatory \citep{Schodel_2002, Ghez_2003, Ghez_2008, Gillessen_2017}. Since 2016, the GRAVITY instrument at the Very Large Telescope Interferometer (VLTI), which coherently combines the four 8-meter class telescopes of the observatory, has provided astrometric measurements with unprecedented precision, achieving accuracies as fine as 30 \text{$\mu as$}, more than 15 times better than what is possible with a single 8-meter telescope \citep{GRAVITY_2017}.

This has enabled highly accurate measurements of the SMBH mass, $m_\bullet = 4.3 \times 10^6 \, \text{M}_\odot$, with a precision of $0.3\%$, and its distance from us, $R_0 = 8.3 \,$ kpc, with a precision of $0.1\%$. Observations of the pericenter passage of S2 in 2018, when the star reached a distance of $1400$ Schwarzschild radii ($R_s$) from Sgr A* with a speed of about $7700$ km s$^{-1}$ $\simeq$ $0.03 \,$ c, allowed the first direct tests of General Relativity (GR) in the vicinity of an SMBH \citep{GRAVITY_2018,GRAVITY_2020}. In particular, the first-order effects in the post-Newtonian (PN) expansion of GR on its orbital motion were detected, which are the gravitational redshift of spectral lines and the prograde in-plane precession of the orbit of $\delta \varphi = 3 \pi R_S/[a(1-e^2)]\approx 12$ arcmin per orbit, where $a$ and $e$ are the semi-major axis and eccentricity.

In order to rigorously test whether the orbit of S2 follows a Schwarzschild orbit as predicted by GR at 1PN order, in \citet{GRAVITY_2020, GRAVITY_2022, GRAVITY_2024} the acceleration of the star is modeled using a 1PN approximation for a massless test particle \citep{Will_1993}. The 1PN terms are multiplied by a factor $f_{SP}$, such that $f_{SP}=0$ represents a Newtonian orbit and $f_{SP}=1$ represents a Schwarzschild, GR, orbit. Using data up to September of 2022, in \citet{GRAVITY_2024} it is found $f_{SP}=1.135 \pm 0.110$, indicating that a Schwarzschild orbit is strongly favored over a Newtonian orbit at a $10 \sigma$ confidence level.
In addition, in \cite{GRAVITY_2024} constraints are given on the amount of the extended mass that could be distributed around Sgr A* that, if spherically symmetric around the massive object, would induce a retrograde precession of the S2 orbit. The mentioned analysis sets an upper limit of $\approx 1200 \, \text{M}_\odot$ to the extended mass contained within the orbit of S2, namely within the central $\approx 0.01$ pc of the Galaxy. This value aligns with the predictions from numerical simulations by \citet{Zhang_Pau_2023} for a dynamically relaxed stellar cusp surrounding Sgr A*, as composed of old stars and stellar remnants, including stellar-mass black holes. 

The existence of such a stellar cusp around SMBHs in galactic centers is a long‐standing prediction of stellar dynamics.  Over a timescale of the order of the two‐body relaxation time, a stellar population near an SMBH is expected to settle into an equilibrium distribution well inside the SMBH’s influence radius. To a good approximation, the resulting density profile follows a power law, as first shown by \citet{Peebles_1972}, and later by \citet{frank_rees_1976} and \citet{BW_1976_1}. For a multi‐mass stellar population, \cite{BW_1977_2} provided a heuristic solution based on the (unphysical) assumption that the mass was equally divided between light stars and heavier stellar‐mass black holes.  A more physical, self‐consistent, solution for a general stellar population was later derived by \cite{Tal_2009} and \cite{Preto_Pau_2010}, who showed how mass segregation drives heavier objects such as stellar-mass black holes into steeper cusps, thus dominating the mass distribution in the innermost regions, namely within S2’s apocenter of \(\sim0.01\) pc for the case of the GC.
These compact objects are  potential sources of extreme-mass-ratio inspirals (EMRIs), that will likely be observed by the future LISA mission \citep{LISA_2017, Pau2018LRR}. 
Numerical simulations, including those by \citet{Freitag_2006}, \citet{Tal_2009}, and more recently by \citet{Zhang_Pau_2023}, suggest that up to 100 stellar-mass black holes could reside within the apocenter of S2 in the GC. This population corresponds indeed to a granular rather than smooth mass distribution, potentially perturbing the orbit of S2. The possibility that an intermediate-mass black hole (IMBH) companion to Sgr A* resides within the orbit of S2 has also been explored, showing that it can only have a mass $<1000 \, \text{M}_\odot$ in order to be compatible with observations \citep{GRAVITY_2023, Will_2023}.

\citet{Merritt_2010} and \citet{Sabha_2012} first investigated the impact of perturbations caused by a cluster of stellar-mass black holes on stellar orbits near Sgr A*. \citet{Merritt_2010} found that such perturbations could introduce a precession of the orbital plane comparable in magnitude to that caused by the Lense -- Thirring effect, potentially complicating the detection of the spin and quadrupole moment of Sgr A*. They concluded that detecting the spin of Sgr A* would require observation of stars with semi-major axes $\lesssim 1$ mpc, where relativistic effects would dominate over perturbations. \citet{Zhang_Iorio_2017} later confirmed this result, analyzing the impact of these perturbations on the orbital motion and redshift of the stars. However, they pointed out that, in principle, the dynamical signatures of stellar perturbations might be distinguishable from GR spin effects, suggesting that the conclusions of \citet{Merritt_2010} could be somewhat pessimistic.

In this work, we conduct an extensive statistical study to assess the impact of the \textit{granularity} of the mass distribution in the GC on the orbit of the S2 star, as compared to a smooth, spherically symmetric distribution. 

Since the spin of Sgr A* produces negligible (at the level of present observational facilities) effects on the motion of S2, we consider it sufficient to model Sgr A* as a Schwarzschild SMBH. 
Specifically, we simulate the orbit of S2 around a Schwarzschild SMBH representing Sgr A* as surrounded by a star cluster composed of equal-mass objects.
To study the effects of this granular mass distribution, in Section \ref{sec:simp_approach} we report results coming from a large set of numerical simulations performed using a fast, simplified dynamical approach, varying both the total mass of the cluster and the mass of individual cluster components. In Section \ref{sec:comp_nbody} we compare a subset of these results to those obtained using some full $N$-body simulations.
Given the remarkable precision now achievable with GRAVITY, in Section \ref{sec:fitting} we perform a mock data analysis to investigate whether the granularity of the mass distribution could cause observable deviations in the motion of S2 with respect to its best-fit Schwarzschild orbit. Finally, in Section \ref{concl}, we draw our conclusions and discuss the implications of our findings for future observations.


\section{Simplified dynamical approach}
\label{sec:simp_approach}

\subsection{Method}
\label{metod1}
The aim of this paper is to study the relevance of the granularity of the gravitational field acting on the S2 star orbiting Sgr A$^*$. To do this, 
we simulate a test star (S2) orbiting an SMBH of mass $m_\bullet$ (Sgr A$^*$), using our own quadruple-precision Fortran 90 integrator. It is a one-body code which integrates the equations of motion of the test star by means of a 4th-order Runge-Kutta method with variable time stepping. 
The star starts its motion from the apocenter in the $x$-$y$ plane, with initial conditions:
\begin{equation}
\label{ic}
    \mathbf{r}_{0} =(r_{a,S2}, 0, 0), 
~\mathbf{v}_{0} = (0, -v_{a,S2}, 0).
\end{equation}
At zeroth order (Newtonian case) the apocenter distance is $r_{a,S2}=a(1+e)$ and the speed at apocenter is $v_{a,S2}= \sqrt{2/r_{a,S2} - 1/a}$, where $a$ and $e$ are, respectively, the semimajor axis and eccentricity of the unperturbed Newtonian orbit.

We assume that a number $N$ of point-mass particles of equal mass $m$ are distributed around the central SMBH. We sample their positions from a power-law density distribution $\rho(r) = \rho_0 ({r}/{r_0})^\alpha$, with $\alpha<0$, which is approximately the steady-state density distribution of a cluster of stars and stellar remnants around an SMBH.
We note that the two‐body relaxation time evaluated at 
$r = r_{a,\mathrm{S2}}\approx0.01\ \mathrm{pc}$
for a stellar cusp of identical‐mass objects is (following \citealp{Pau2018LRR}, eq.~15) $t_{rlx} \gtrsim 5\times10^7\ \mathrm{yr}$ for $m \leq 100\,M_\odot$, far exceeding the 16 yr orbital period of S2.  Consequently, the radial density profile does not change appreciably over a single S2 orbit.

We sample the positions of the $N$ objects up to a cut radius $r_{cut} > r_{a,S2}$. We choose $\alpha=-2$ and $r_{cut}=2r_{a,S2}$, upon verification that the results depend weakly on the choice of $\alpha$ and $r_{cut}$, provided the enclosed mass within the apocenter of S2 is kept constant and $\alpha$ is within reasonable limits predicted for a stellar cusp (see for example \citealt{BW_1977_2, Tal_2009}).
The position sampling is done by inverting the cumulative mass distribution, defined as $M(<r)/M_t$, where $M_t=Nm$ is the total cluster mass. We assume that each body is fixed in space, with position $\mathbf{r}_j(t)= \mathbf{r}_{j0}$ and velocity $\mathbf{v}_j(t)=0$, $\forall \, t$ and $j=1,..,N$.

For a reference frame centered on the SMBH, the equation of motion for the test star is:

\begin{equation}
\label{eq_NBsystem}
    \ddot{\mathbf{r}} =- G m_\bullet \frac{\mathbf{r}}{r^3} +
    \mathbf{f}_{{1PN}} + G \sum_{\substack{j=1 }}^N m\frac{\mathbf{r}_{j} - \mathbf{r}_{S2}}{|\mathbf{r}_{j} - \mathbf{r}_{S2}|^3}
\end{equation}
where $G$ is the gravitational constant and $\mathbf{f}_{{1PN}}$ is the first-order PN force per unit mass acting on the test star due to the presence of the SMBH \citep{mowill_2004}. We neglect here higher order PN terms, such as those accounting for the spin of a Kerr, rotating SMBH, as they give a negligible contribution to the orbital motion of S2 \citep{Cap_Sad_2023}. 


\begin{figure*}[h!]
\centering
\begin{tikzpicture}[
    node distance=0.8cm and 0.8cm,
    every node/.style={draw, align=center, rounded corners, minimum width=1.5cm, minimum height=0.8cm, font=\small},
    scenario/.style={fill=green!15},
    parameter/.style={fill=blue!15},
    explanation/.style={align=center, font=\small\itshape, draw, rounded corners, minimum width=2.0cm, minimum height=0.8cm, fill=gray!20},
    arrow/.style={-stealth, thick, draw=black}
]

\node[draw=none] (plot) at (-8, 0) {
    \includegraphics[width=0.55\textwidth]{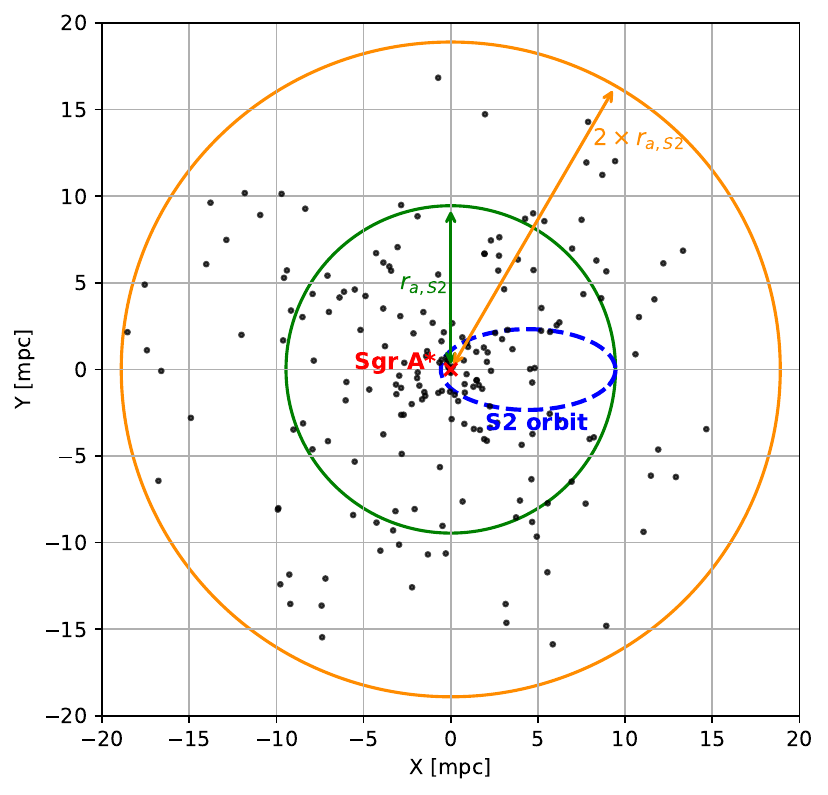}
};

\node[scenario] (scenario1) at (2, 3) {Loop over total mass in particles within the apocenter of S2: \\ $M_{e,S2} = 100, 500, 1000, 1500 \, \text{M}_\odot$. \\ Loop over the particle mass: \\ $m = 1, 2, 5, 10, 20, 50, 100 \, \text{M}_\odot$.};

\node[explanation] (explanation1) [below=1cm of scenario1] {100 different samplings of the positions of the particles, \\ according to $\rho(r) \propto r^{-2}$. \\ Particles are assumed to be fixed in space.};


\node[parameter] (explanation2) [below=1cm of explanation1] {Simulation of the motion of S2, \\ for each sampling of the particle positions.};

\draw[arrow] (scenario1.south) -- (explanation1.north);
\draw[arrow] (explanation1.south) -- (explanation2.north);

\end{tikzpicture}
\caption{Left: The S2 orbit (in blue) around Sgr A* (marked by a red cross) and a particular realization (black dots) of the distribution of the surrounding cluster particles, as described in Section \ref{metod1}. The total mass in cluster particles enclosed within the apocenter distance of S2, $r_{a,S2}$ (in green), is denoted as $M_{e,S2}$. The particles are sampled up to a distance $2 \times r_{a,S2}$ (in orange). Right: Flowchart describing the analysis procedure.}
\label{fig:simulation-scenarios}
\end{figure*}
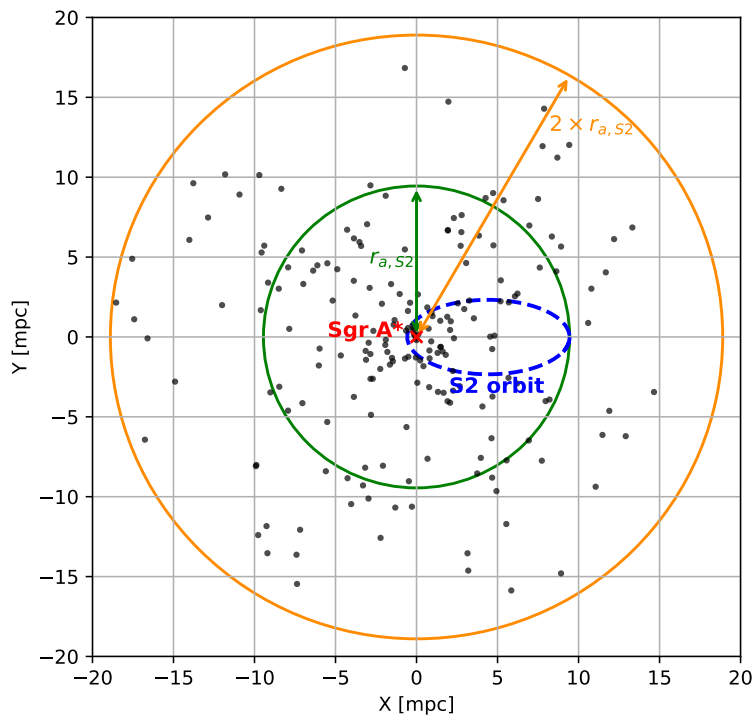

We thus model the orbit of the test star as the solution of a one-body problem, where the star moves under the gravitational influence of the SMBH and the $N$ bodies of the cluster. 
We neglect mutual interactions among the cluster bodies and with the SMBH, treating them as fixed particles in space that provide a \textit{granular} contribution to the otherwise smooth gravitational field.
This approach is significantly faster than time integration with a full $N$-body code, as its execution time scales as  $\mathcal{O}(N)$ instead of  $\mathcal{O}(N^2)$. This efficiency enables a large statistical study that would be otherwise infeasible with a full $N$-body approach.

We consider different cluster and particle masses,
 as illustrated in Figure~\ref{fig:simulation-scenarios}. The enclosed mass within the apocenter of S2, $M_{e,S2}$, varies from $100 \, \text{M}_\odot$ to $1500 \, \text{M}_\odot$, consistent with the upper limit of approximately $1200 \, \text{M}_\odot$ at the $1 \sigma$ confidence level reported by \citet{GRAVITY_2024}. The enclosed mass is distributed among cluster particles of equal mass $m$, which varies from $1 \, \text{M}_\odot$ to $100 \, \text{M}_\odot$ (with the number $N$ of cluster particles changing accordingly). We thus explore cases ranging from a cluster of 1 M$_\odot$ stellar objects to a cluster of stellar mass black holes with masses up to 100 M$_\odot$, motivated by the observed population of merging binary black holes \citep{Abbot2023b}.

{

For each choice of $M_{e,S2}$ and $m$, we performed 100 simulations, each corresponding to a different realization of the spatial distribution of the cluster objects, in order to have an adequate statistic. Each simulation is carried out over 1.1 times the radial period of S2.

The values of the semi-major axis and eccentricity of S2, and of the mass of Sgr A*, are taken from Table B.1 of \cite{GRAVITY_2022}. 
For computational convenience, the units are chosen such that $G=1$, the unit of mass is the SMBH mass $m_\bullet$, and the unit of length is $D= 10$ mpc, which corresponds roughly to the apocenter distance of S2. 


\subsection{Results}
\label{res:simp_approach}
The first thing to note in our numerical simulations is that the granularity of the potential naturally leads to a breaking of spherical symmetry. The departure from spherical symmetry of the force acting on the test star can be quantified by $\Delta f = |\mathbf{f}\cdot \hat{\mathbf{r}}|/f-1$, where $\mathbf{f}$ is the total force per unit mass acting on S2 at any given time, $f$ is its magnitude, and $\hat{\mathbf{r}}$ is the radial unit vector. This quantity is zero for a spherically symmetric force field. However, as illustrated in Fig.~\ref{fig_scprod}, we find deviations from zero. 
These deviations, in principle, arise from a combination of global deviations from spherical symmetry, and local scattering events between the star and the cluster objects.
To investigate the main cause of these deviations, we evaluate the ratio 
$F = \sum_{i=1}^{N_S} f_{i}/{f}_{SgrA*}$ between the norm of the force due to the interaction of the test star with the cluster objects and that of the force due to Sgr A$^*$, considering only the number $N_S<N$ of encounters giving $f_{i}/{f}_{SgrA*} > 10^{-4}$. 
The overlap between $F$ and $\Delta f$ in Fig.~\ref{fig_scprod} clearly indicates that deviations from spherical symmetry in the simulations are due to the strongest encounters occurring between S2 and the cluster objects.
Depending on the specific sampling of the positions of the cluster objects, the strength of the scattering interactions varies, with the force ratio $f_{i}/{f}_{SgrA*}$ reaching $\sim 10^{-2}$ for the strongest interactions observed across the simulations performed. 
This shows the great importance of performing a large statistical study in order to correctly analyze the effect of the granularity of the mass distribution around the central SMBH on the orbit of S2. To obtain significant results we need a large set of simulations with different realizations of the sampling of the cluster objects.

\begin{figure}[h!]
    \centering
    \includegraphics[width=0.48\textwidth]{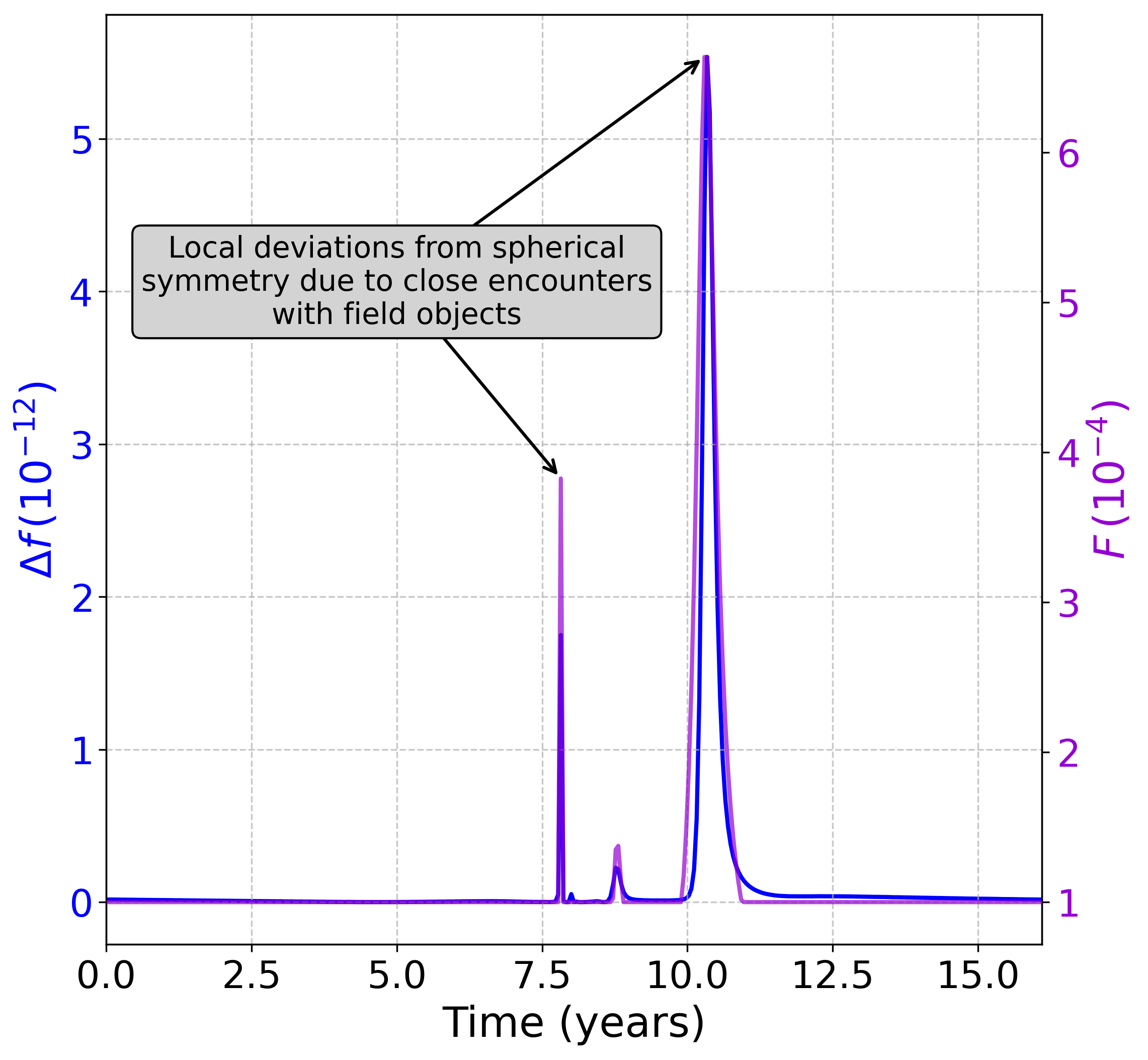}
     \caption{
     In blue: deviation from spherical symmetry as a function of time, quantified by $\Delta f$ (see text in Section \ref{res:simp_approach}), for one arbitrarily chosen  simulation of the S2 orbit in presence of a cluster of $200$ objects of $10 \, \text{M}_\odot$ each. In purple: the force ratio $F$ (see text in Section \ref{res:simp_approach}).
     }
    \label{fig_scprod}%
\end{figure}

The orbit of a test star around a Schwarzschild SMBH is planar and characterized by an in-plane, prograde precession of the pericenter angle. The introduction of a smooth, spherically symmetric mass distribution surrounding the SMBH maintains the planarity of the orbit of the star and adds a retrograde precession. In these cases, both the orbital energy and angular momentum are conserved quantities.
Breaking the spherical symmetry of the potential, as in the granular case, implies that the orbit of the star is no longer planar. In our case, the star and its velocity initially lie in the $x-y$ plane, so that spherical symmetry breaking implies that the star acquires a $z$ displacement and a $v_z$ velocity component.
While the orbital energy is still conserved, as our system is static,  neither the modulus nor the direction of the angular momentum are conserved anymore. Consequently, in addition to the in-plane orbital precession, a precession of the instantaneous orbital plane occurs.


The precession angle $\theta$ of the orbital plane with respect to the initial $x-y$ plane is calculated  as 
\begin{equation}
\label{alpha}
    \theta(t) = \text{arccos} \left( \frac{\vec{L}(t) \cdot \vec{L}(0)}{L(t) L(0)} \right),
\end{equation}
where $\vec{L}(t)$ is the orbital angular momentum at time $t$. 

The in-plane precession can be calculated by considering two consecutive apocenter passages of the test star, such that after a full radial period the angular shift is equal to 
\begin{equation}
    \label{phi}
    \delta \varphi_{xy} = \text{arctan} \left( \frac{\Delta y _{\text{apo}}}{\Delta x_{\text{apo}}} \right).
\end{equation}
This quantity corresponds exactly to the in-plane orbital precession only if the test star, starting its motion in the $x-y$ plane, would follow a planar orbit. Since in our simulations the orbit is non-planar, this quantity represents the angular precession of the orbit projected onto the $x-y$ plane, which is the initial orbital plane of the test star.

\begin{figure*}[h!]
        \centering
        \captionsetup[subfigure]{position=top} 
        \begin{subfigure}[b]{0.48\textwidth}
            \centering
            \includegraphics[width=\textwidth]{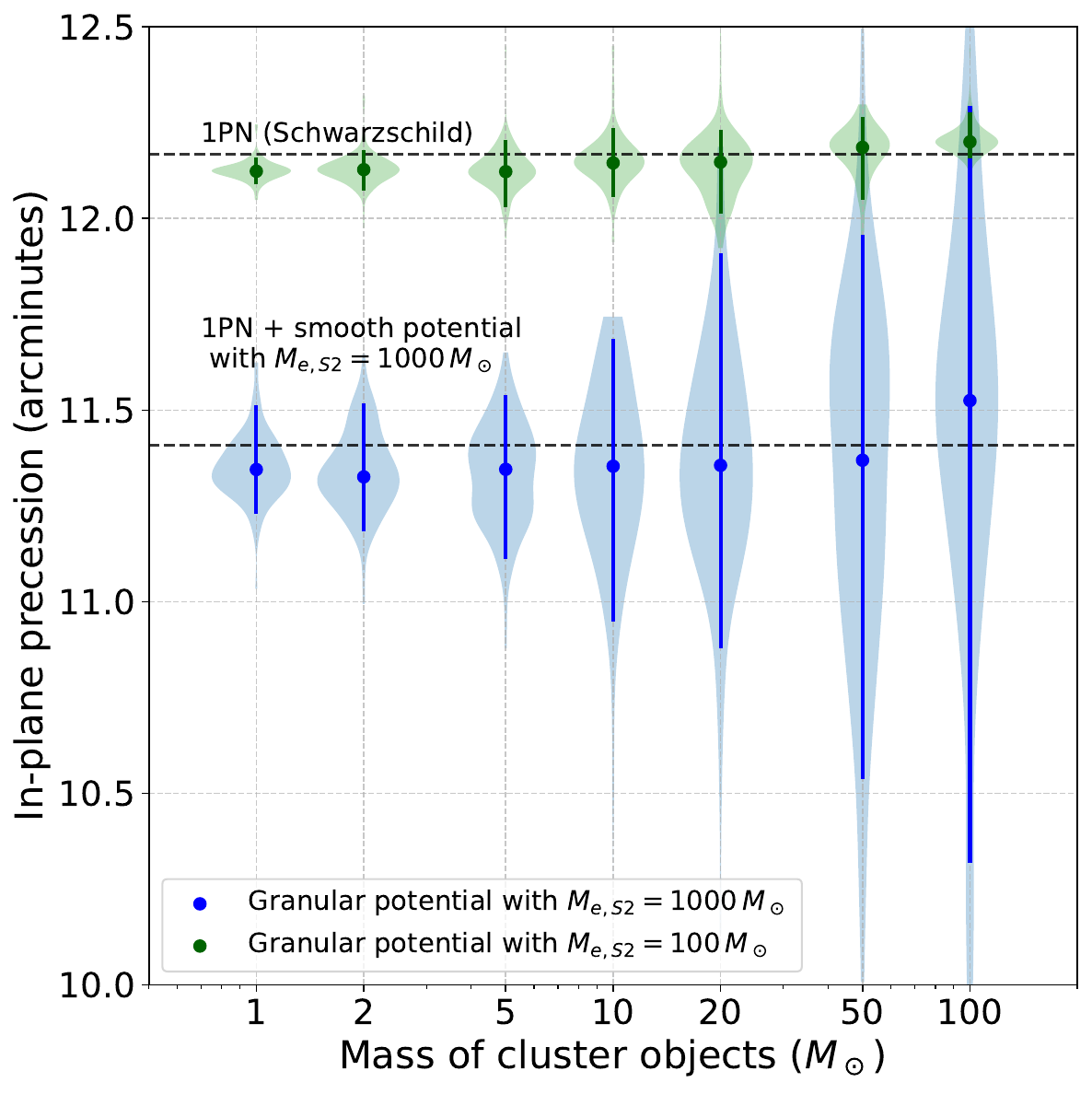}
        \end{subfigure}
        \hfill
        \begin{subfigure}[b]{0.48\textwidth}  
            \centering 
            \includegraphics[width=\textwidth]{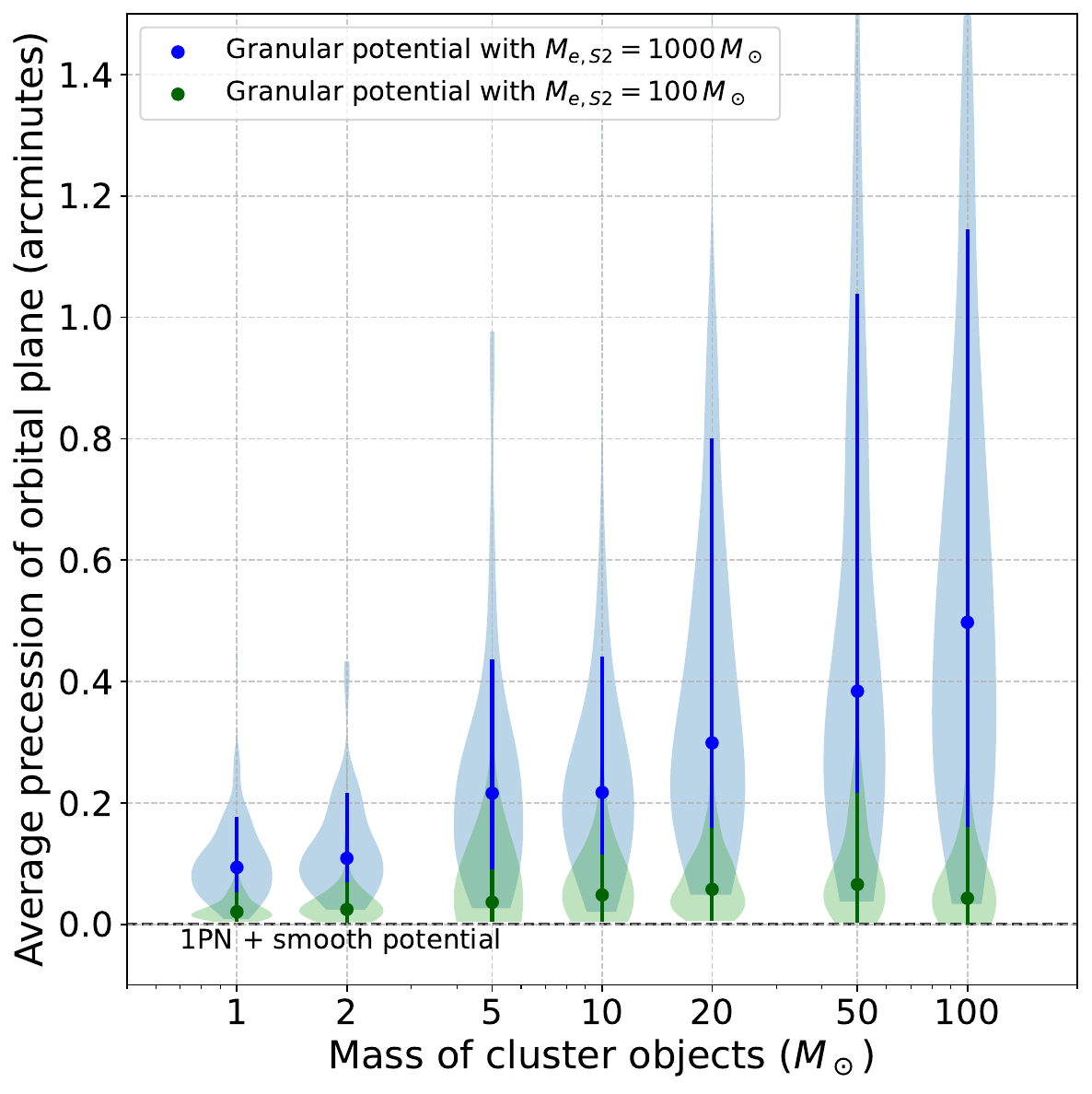}
        \end{subfigure}
        \hfill
        \caption[]
        {Violin plots showing the in-plane angular precession (left panel) and the average orbital plane precession (right panel) as a function of the mass of the cluster objects, for $M_{e,S2}=100 \, \text{M}_\odot$ (green) and $M_{e,S2}=1000 \, \text{M}_\odot$ (blue).
        The dots show the median value of the distributions, the vertical segments indicate the breadth of the distributions, going from the 5th to the 95th percentile in case of the in-plane precession and from the minimum value to the 90th percentile in case of the orbital plane precession.
        }
        \label{fig:plots_fixed}
\end{figure*}

As described in Section \ref{metod1} and summarized in Figure~\ref{fig:simulation-scenarios}, we vary both the total mass of the cluster of particles enclosed within the apocenter distance of S2, $M_{e,S2}$, and the mass $m$ of the individual particles. 
For each configuration, we perform 100 simulations of the motion of S2 with different realizations of the positions of the particles, which lead to different trajectories. For each simulation, we compute the in-plane precession, $\delta \varphi_{xy}$, and the average precession of the orbital plane over a full radial period, $\langle \theta \rangle$ (Eqs.~\ref{alpha} and~\ref{phi}). 
We choose 100 sampling realizations as a practical compromise. A larger ensemble would give a more reliable statistical study, however, we checked that the results on $\delta \varphi_{xy}$ and $\langle \theta \rangle$ obtained with our sampling choices present statistical uncertainties subdominant to physical ones.\footnote{The Kullback–Leibler divergence with $N_1=50$ and $N_2=100$ realizations is $D_{\rm KL}(N_1|N_2)\lesssim0.05$.} This ensures reliable statistics without excessive computational cost.

We find that the trajectory of S2 indeed deviates from a planar orbit, with the deviation depending on the specific sampling of the positions of the cluster particles. Across the simulations conducted, we observe that the orbital plane of S2 can precess by up to a significant fraction of the in-plane Schwarzschild precession (12 arcmin per orbit). 
To better visualize and quantify the impact of the granularity of the mass distribution on the orbit of S2, in Figure~\ref{fig:plots_fixed} we present violin plots \citep{Hintze_1998} illustrating the distribution of $\delta \varphi_{xy}$ and $\langle \theta \rangle$ for two representative cases, corresponding to $M_{e,S2}=100 \, \text{M}_\odot$ and $M_{e,S2}=1000 \, \text{M}_\odot$, as a function of the mass of the cluster particles. Violin plots are useful in this case as they show the median, spread, and shape of the $\delta \varphi_{xy}$ and $\langle\theta\rangle$ distributions.

For $M_{e,S2}=100 \, \text{M}_\odot$, the median in-plane precession is close to the value predicted for a 1PN Schwarzschild orbit (the retrograde precession caused by a smooth distribution with $M_{e,S2}=100 \, \text{M}_\odot$ is negligible), with little spread in the distribution. The average orbital plane precession is similarly small, remaining below 0.2 arcmin. This is expected in a regime where the number of cluster particles is very low (100 particles of $1 \, \text{M}_\odot$ or one particle of $100 \, \text{M}_\odot$), reducing the probability of scattering events to occur. Thus, deviations from a Schwarzschild orbit are not very pronounced.

Increasing $M_{e,S2}$, deviations from a smooth potential become more noticeable. 
The median in-plane precession remains consistent with the value predicted in a smooth potential, but the distribution broadens significantly with increasing mass (and correspondingly with decreasing number) of the cluster objects, reflecting the more pronounced granularity of the gravitational potential. This can be explained by the fact that more massive cluster objects act as stronger perturbers, as illustrated in Fig.~\ref{fig_cdf}, inducing larger trajectory deviations with respect to an orbit in a smooth potential.
For $M_{e,S2}=1000 \, \text{M}_\odot$, the width of the in-plane precession distribution (5th–95th percentile) increases from $\approx 0.2$ arcmin for $m=1 \, \text{M}_\odot$ to $\approx 2$ arcmin for $m=100 \, \text{M}_\odot$ (see Figure~\ref{fig:plots_fixed}). We find a deviation with respect to the value expected in a smooth potential up to $\approx 1.5$ arcmin, corresponding to a fractional variation of $\approx 13\%$. The average orbital plane precession also grows with increasing $m$ (decreasing $N$), with a median value of $\approx 0.1$ arcmin for $m=1 \, \text{M}_\odot$ and $\approx 0.5$ arcmin for $m=100 \, \text{M}_\odot$. The distribution also becomes broader with increasing $m$, reaching up to $\approx 1.2$ arcmin (90th percentile), namely $\approx 10\%$ of the Schwarzschild in-plane precession.

Further increase of $M_{e,S2}$ causes a broader in-plane precession distribution for each value of $m$, and the orbital plane precession distribution has even larger median and range. 
In Table \ref{tab_results} we give the median and range of both distributions for each value of $M_{e,S2}$ and $m$ considered.
Overall, we find that deviations from a smooth potential are larger in scenarios where the cluster is divided into more massive (and fewer) particles, in line with intuitive expectations, as the granular fluctuations over the mean field become more pronounced and the interactions between S2 and the cluster objects are stronger on average. 
In addition, such deviations become more pronounced with increasing total enclosed mass $M_{e,S2}$, as the number of scattering events increases due to the larger number of cluster particles. These results indicate that the granularity of the mass distribution plays a significant role in shaping the S2 orbital properties, so that approximating the actual mass distribution in this region with a smooth potential could be highly inaccurate.

\begin{table*}[h!]
    \centering
    \caption{Median values of $\delta \varphi_{xy}$ and $\langle \theta \rangle$ (in arcminutes) obtained for different choices of $M_{e,S2}$ and $m$. Uncertainties give the range of the distributions, going from the 5th to the 95th percentiles for $\delta \varphi_{xy}$ and from the minimum to the 90th percentile for $\langle \theta \rangle$.}
    \label{tab:precession_values}
    \renewcommand{\arraystretch}{1.2}
    \begin{tabular}{|c|c|c|c|c|}
        \hline
        \hline
        & \(M_{e,S2} = 100 M_\odot\) & \(M_{e,S2} = 500 M_\odot\) & \(M_{e,S2} = 1000 M_\odot\) & \(M_{e,S2} = 1500 M_\odot\) \\
        \hline
        \(m = 1 M_\odot\)   & \(12.12_{-0.03}^{+0.03}, 0.021_{-0.017}^{+0.032}\) & \(11.78_{-0.09}^{+0.06}, 0.062_{-0.050}^{+0.059}\) & \(11.35_{-0.12}^{+0.17}, 0.094_{-0.085}^{+0.084}\) & \( - \) \\
        \(m = 2 M_\odot\)   & \(12.13_{-0.05}^{+0.05}, 0.024_{-0.021}^{+0.045}\) & \(11.80_{-0.13}^{+0.10}, 0.064_{-0.052}^{+0.102}\) & \(11.33_{-0.14}^{+0.19}, 0.109_{-0.085}^{+0.107}\) & \(10.91_{-0.24}^{+0.22}, 0.160_{-0.125}^{+0.243}\) \\
        \(m = 5 M_\odot\)   & \(12.12_{-0.09}^{+0.08}, 0.036_{-0.032}^{+0.054}\) & \(11.78_{-0.22}^{+0.17}, 0.112_{-0.093}^{+0.171}\) & \(11.35_{-0.23}^{+0.19}, 0.216_{-0.190}^{+0.220}\) & \(10.94_{-0.29}^{+0.33}, 0.245_{-0.214}^{+0.287}\) \\
        \(m = 10 M_\odot\)  & \(12.15_{-0.09}^{+0.09}, 0.048_{-0.044}^{+0.066}\) & \(11.77_{-0.31}^{+0.29}, 0.138_{-0.126}^{+0.251}\) & \(11.35_{-0.40}^{+0.33}, 0.217_{-0.197}^{+0.224}\) & \(10.84_{-0.45}^{+0.39}, 0.263_{-0.201}^{+0.358}\) \\
        \(m = 20 M_\odot\)  & \(12.15_{-0.14}^{+0.08}, 0.057_{-0.052}^{+0.101}\) & \(11.79_{-0.37}^{+0.37}, 0.217_{-0.184}^{+0.231}\) & \(11.36_{-0.48}^{+0.55}, 0.299_{-0.250}^{+0.502}\) & \(10.91_{-0.60}^{+0.51}, 0.357_{-0.291}^{+0.477}\) \\
        \(m = 50 M_\odot\)  & \(12.19_{-0.14}^{+0.08}, 0.066_{-0.063}^{+0.150}\) & \(11.84_{-0.62}^{+0.38}, 0.287_{-0.266}^{+0.558}\) & \(11.37_{-0.83}^{+0.59}, 0.384_{-0.346}^{+0.655}\) & \(10.92_{-0.94}^{+0.76}, 0.484_{-0.434}^{+0.846}\) \\
        \(m = 100 M_\odot\) & \(12.20_{-0.04}^{+0.08}, 0.043_{-0.043}^{+0.117}\) & \(11.91_{-0.69}^{+0.53}, 0.311_{-0.301}^{+0.495}\) & \(11.53_{-1.21}^{+0.77}, 0.497_{-0.464}^{+0.647}\) & \(10.96_{-1.08}^{+1.14}, 0.575_{-0.464}^{+1.358}\) \\
        \hline
    \end{tabular}
    \label{tab_results}
\end{table*}

To conclude, due to mass segregation, we expect that the mass distribution within the S2 orbit (in the central 0.01 parsec) is dominated by stellar mass black holes \citep{GRAVITY_2024}. Approximating the mass distribution in this region with a population of light stars with masses of 1 M$_\odot$ and heavier stellar black holes with masses of 10 M$_\odot$, \citet{GRAVITY_2024} predicts that around 1200 M$_\odot$ of extended mass lies within the apocenter of S2, of which $2/3$ (800 M$_\odot$) is in stellar black holes and $1/3$ (400 M$_\odot$) is in stars. We perform 100 simulations in this scenario and confirm that the stellar black holes dominate the scattering, being the main contributors to deviations of the actual S2 orbit from the one in a smooth potential. In this scenario, they cause an average orbital plane precession of up to $\approx$ 0.5 arcmin and a fractional variation of the in-plane precession of up to $\approx$ 4.5\%. Perturbations from the stars are instead negligible.

\begin{figure}[h!]
    \centering
    \includegraphics[width=0.48\textwidth]{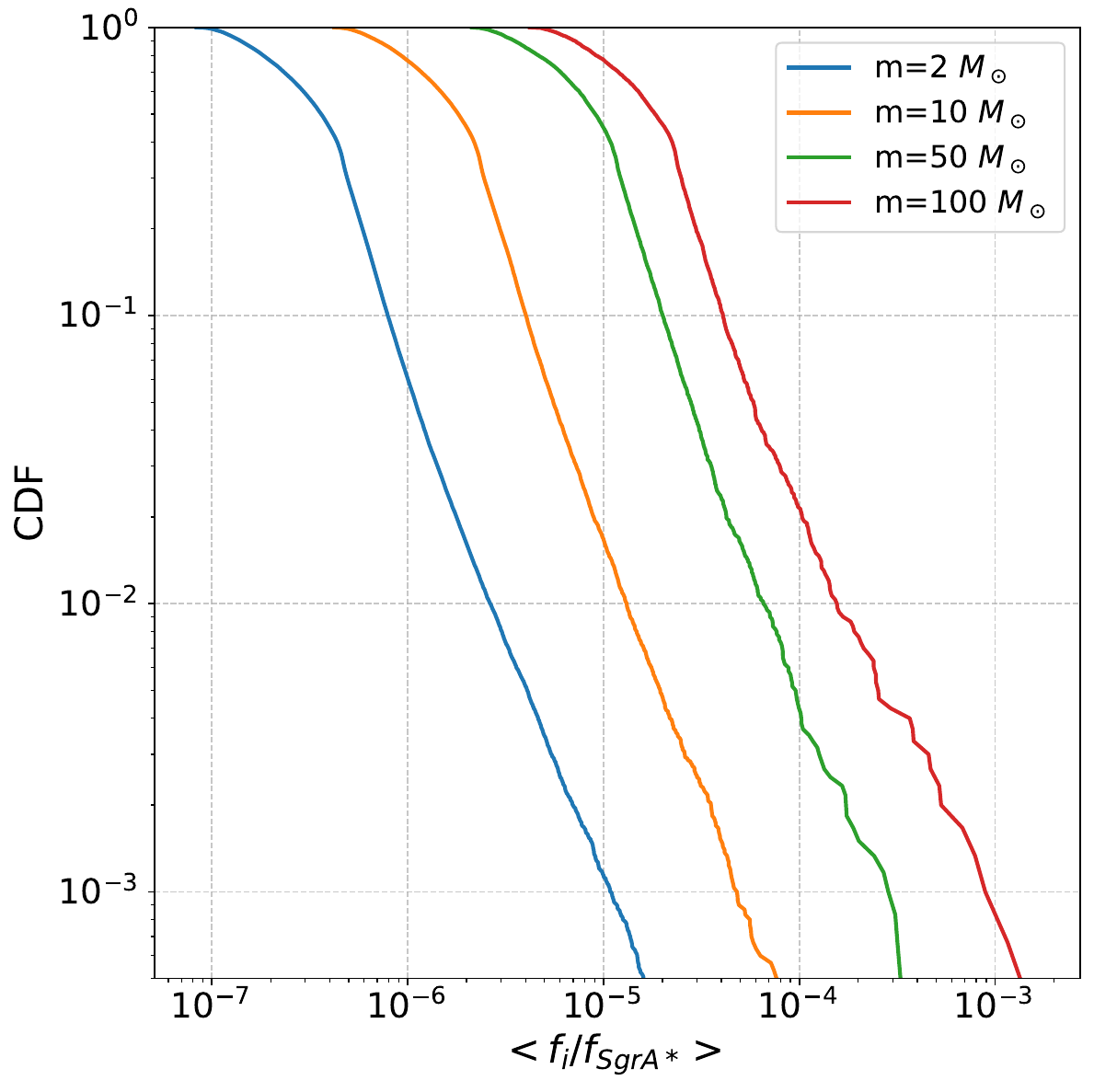}
     \caption{
     Cumulative distribution functions (CDF) of the average (over a radial period) of the ratio $f_i/f_{SgrA*}$, between the norm of the force exerted on S2 by each cluster object and that due to Sgr A*. Here $M_{e,S2}$ is set to $1500 \, \text{M}_\odot$ and $m$ is varied as indicated in the figure. Larger $m$ leads to stronger interactions on average between S2 and the cluster particles. 
     }
    \label{fig_cdf}%
\end{figure}

\section{Comparison with full $N$-body approach}
\label{sec:comp_nbody}

In order to validate our simplified dynamical approach, we simulated a subset of the above cases with full $N$-body simulations, 
using the \arwv code
\citep{miktan99b,miktan99a,chacd19,chacd21}, which is a chain-regularized code including post-Newtonian corrections up to 2.5 PN order in the potential exerted by the SMBH on the test, S2, star.

We neglect post-Newtonian cross‐terms (coupling the central SMBH, S2, and the cluster objects), which would arise at 1PN order \citep{Will_2014, AmaroSeoane2025b}, because over a single 16 yr revolution of S2, Newtonian perturbations from the cluster objects dominate any PN coupling effects.
We adopted the same code units as in Section \ref{metod1}.

A key advantage of our simplified approach is its low computational cost. For instance, simulating a single realization of a system with 128 cluster objects of 20 M$_\odot$ 
takes approximately 349 seconds with \arwv on a laptop\footnote{Macbook Pro M1, using a single core}. 
The same simulation using our simplified method requires only 3 seconds on the same machine — a speedup of nearly 120 times. For 100 simulations, this results in a total runtime of about 10 hours with \arwv, compared to just 5 minutes with our simplified method. Additionally, while the computational complexity of \arwv scales as $\mathcal{O}(N^2)$, our approach scales as $\mathcal{O}(N)$, making the difference in execution time even more pronounced as the number of cluster objects increases. Furthermore, \arwv imposes practical limitations on the maximum number of particles that can be simulated. These factors justify our choice to use our simplified approach for the large statistical study in Section \ref{sec:simp_approach}, where we explored a broad range of total cluster masses and individual object masses.

Here, due to the high computational cost associated with $N$-body simulations, we focus on three cases corresponding to cluster objects with masses of 20, 50 and 100 M$_\odot$, such that on average the enclosed mass within the apocenter of S2 is $M_{e,S2}=1000 \, \text{M}_\odot$. For each case, we perform 100 simulations with different samplings of the initial positions and velocities of the cluster objects, as we now describe.


We assume spherical symmetry, and randomly sample the initial orbital elements of the bodies, assuming a uniform semi-major axis distribution, a thermal eccentricity distribution, and isotropic angular coordinates. 
The semi-major axes are sampled from 0 to $10 \, r_{a,S2}$, and particles not passing within $2  \, r_{a,S2}$ during the integration time (1.1 radial periods of S2) are excluded. This process ensures that approximately $N \pm \sqrt{N}$ particles are within the apocenter of S2 at any given time, with some variation as the cluster particles move along their orbits.

To ensure consistency between the two methods under comparison, we use the same initial spatial distribution of the cluster objects, $\vec{r}_j(t=0)$, in both the simplified and full $N$-body simulations. This avoids discrepancies arising from differences in the sampling procedure of the two approaches. Importantly, in the full $N$-body simulations the cluster objects are given non-zero initial velocities $\vec{v}_j(t=0)$, while in the simplified approach the bodies are assumed fixed in space.

To illustrate the difference between the motion of S2 in a fixed potential versus a fully dynamical $N$-body system, we begin with a simple analytical calculation. 
Under the impulse approximation, a close passage at impact parameter \(b\) with relative speed \(v_{\rm rel}\) imparts a transverse kick
\begin{equation}
\Delta v \approx\frac{2\,G\,m}{b\,v_{\rm rel}}.
\end{equation}
For a fixed perturber it is \(v_{\rm rel}=v\), where $v$ is the speed of S2, giving
\(\Delta v_{\rm fixed}=2Gm/(b\,v)\).  

In a full \(N\)-body system the perturber’s own velocity \(v_p\) means
\begin{equation}
v_{\rm rel}=\sqrt{v^2+v_p^2-2\,v\,v_p\cos\alpha},
\end{equation} where $\alpha$ is the angle between the two velocity vectors, 
so \(\Delta v_{\rm N-body}=\Delta v_{\rm fixed}\,(v/v_{\rm rel})\).  
Depending on the encounter geometry \(\alpha\), the kick can be either amplified or suppressed.  Statistically, allowing the perturbers to move, broadens and skews the distributions of changes in velocity \(\Delta v\), and thus in orbital energy and angular momentum (or semi-major axis and eccentricity of the orbit).}

\begin{figure*}[h!]
        \centering
        \captionsetup[subfigure]{position=top} 
        \begin{subfigure}[b]{0.48\textwidth}
            \centering
            \includegraphics[width=\textwidth]{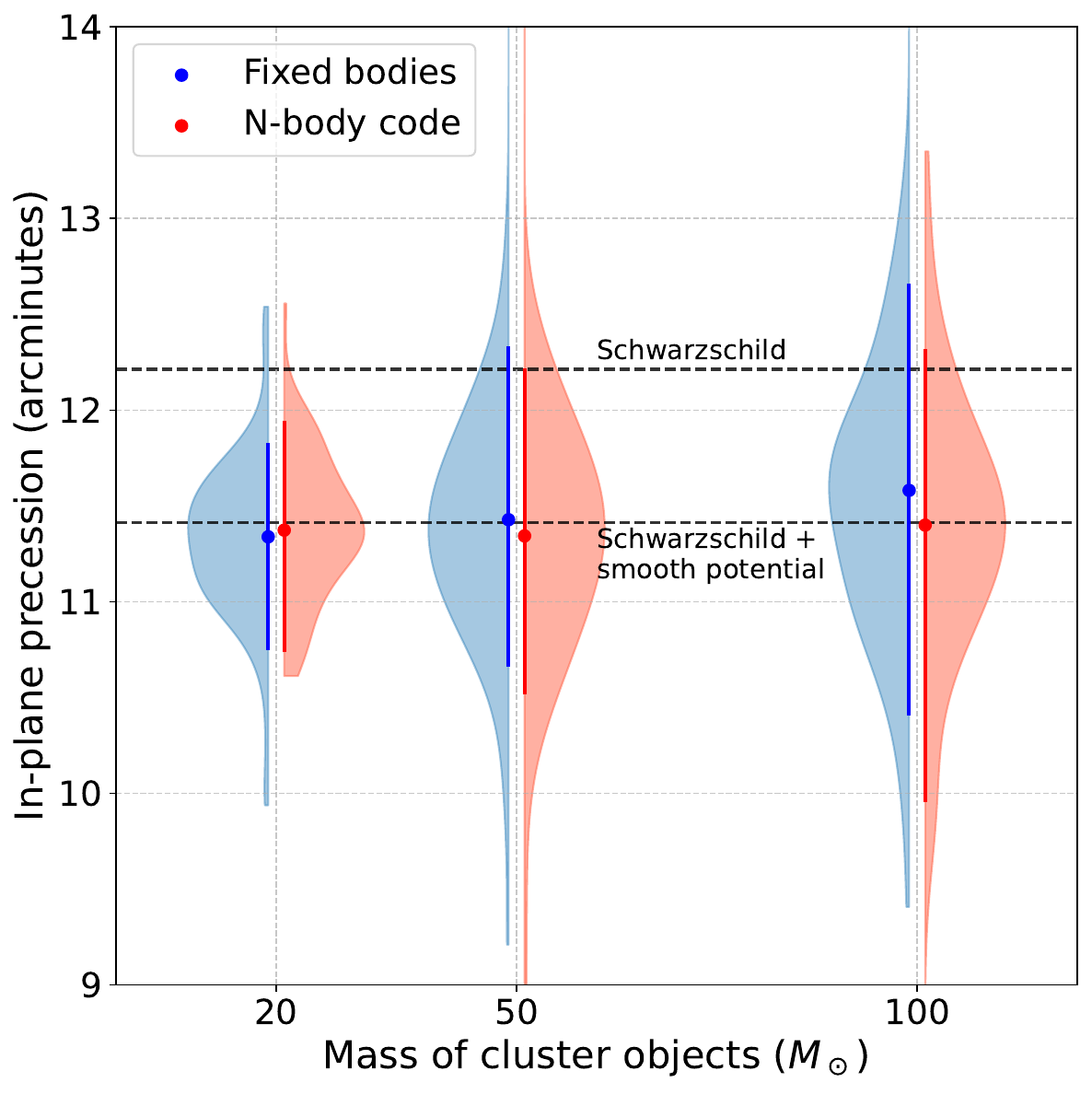}
        \end{subfigure}
        \hfill
        \begin{subfigure}[b]{0.48\textwidth}  
            \centering 
            \includegraphics[width=\textwidth]{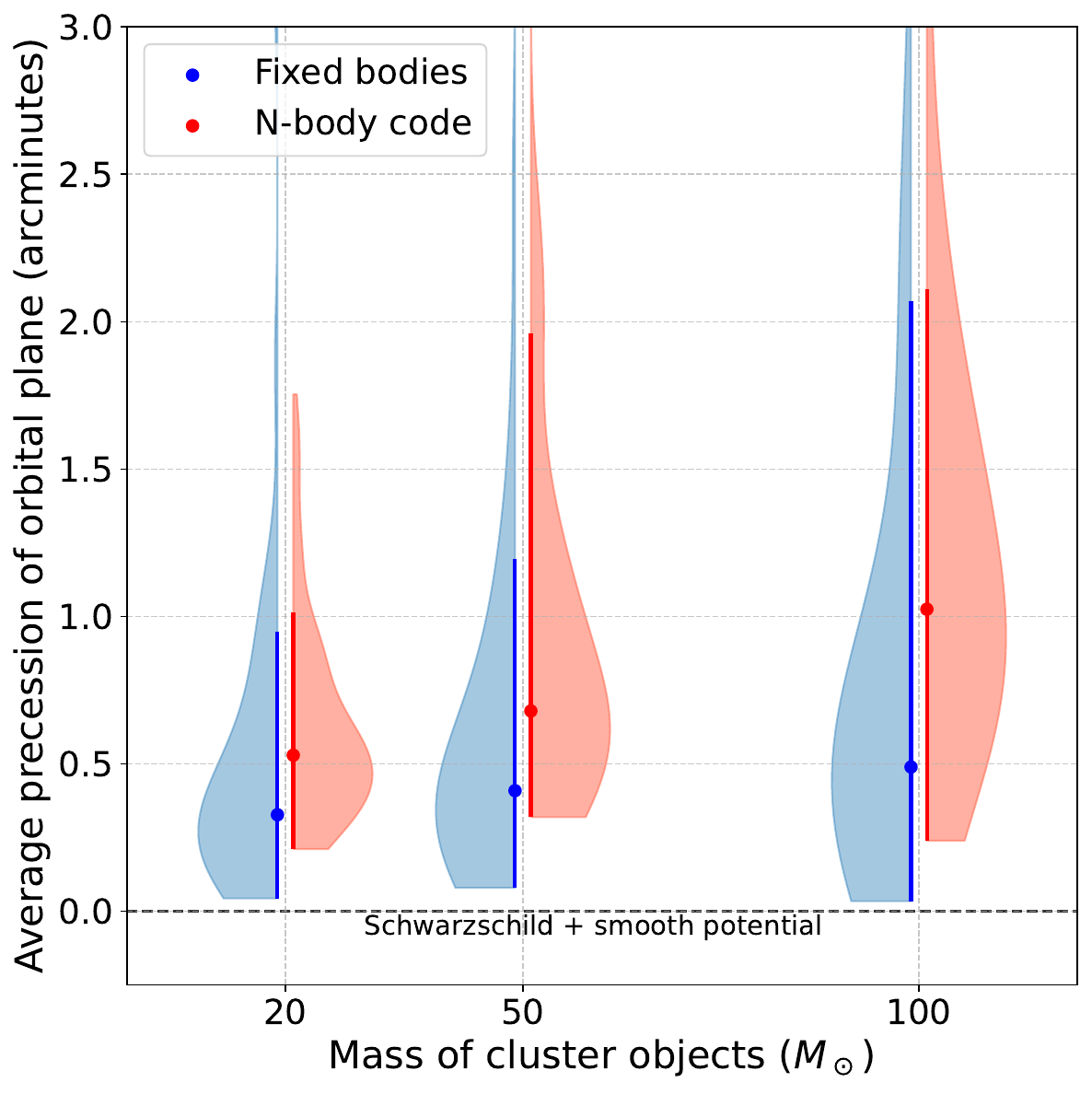}
        \end{subfigure}
        \hfill
        \caption[]
        {Violin plots comparing the results obtained with the simplified approach (in blue) and the full $N$-body code (in red). The plots display the in-plane precession (left panel) and the average precession of the orbital plane (right panel) of S2 over one radial period, as a function of the mass of the cluster objects. The dots show the median value of the distributions, while the vertical segments indicate the breadth of the distributions, going from the 5th to the 95th percentile for the in-plane precession and from the minimum value to the 90th percentile for the orbital plane precession.
        } 
        \label{fig:plots_comp_Nbody}
\end{figure*}

We now compute the same quantities analyzed in Section \ref{res:simp_approach}, namely the in-plane orbital precession and the average orbital plane precession over a full radial period. In the case of full $N$-body simulations (performed in the center of mass frame), we consider the motion of S2 with respect to the central SMBH. A change in the in‐plane precession relative to the smooth‐potential value, and a precession of the orbital plane, both arise from the cumulative effect of many individual kicks over the course of the orbit.
Figure \ref{fig:plots_comp_Nbody} compares the results for the two different methods. The in-plane precession distributions are consistent between the two methods, both in median value and spread, with the median matching the precession expected in case of a smooth potential of the same enclosed mass. However, the orbital plane precession distribution differs: we find a shift toward higher median values in the full $N$-body simulations compared to the simplified approach.
This discrepancy can be explained by the Brownian motion of the central SMBH, which is accounted for in the $N$-body simulations but neglected in the simplified approach. In fact, we confirmed this by using the IAS15 integrator in \reb \citep{rein.liu2012, rein.spiegel2015} and \rebx \citep{tamayo+2020}, which allows us to study the motion of S2 and the cluster bodies relative to a fixed SMBH (i.e. the SMBH is replaced by a fixed potential). We find that, in this case, the median values of the orbital-plane precession distribution are similar to those obtained with the simplified approach. 
However, the general conclusion drawn from the analysis done in Section \ref{sec:simp_approach} is valid also for the full $N$-body simulations: deviations from a smooth potential are larger with more massive (and fewer) cluster particles. 

For completeness, we summarize the mean and standard deviation of the SMBH displacement from its initial position, $\langle \Delta r_{SgrA*} \rangle$, and its velocity, $\langle v_{SgrA*} \rangle$, in the $N$-body simulations for different masses of the cluster objects in Table~\ref{tab:SMBH_motion}.

\begin{table}[h!]
\centering
\caption{Mean and standard deviation of the displacement from the initial position and of the velocity of Sgr A* in the $N$-body simulations for different masses of the cluster objects.}
\label{tab:SMBH_motion}
\begin{tabular}{|c|c|c|}
\hline\hline
$m$ [M$_\odot$] & $\langle \Delta r_{SgrA*} \rangle$ [\text{$\mu as$}] & $\langle v_{SgrA*} \rangle $ [m s$^{-1}$] \\ \hline
20               & $2.84 \pm 1.49$                   & $113.51 \pm 5.0$                \\
50               & $5.49 \pm 1.48$                   & $175.2 \pm 8.3$                \\
100              & $6.49 \pm 3.03$                  & $238.1 \pm 12.7$               \\ \hline
\end{tabular}
\end{table}

These values are consistent with the constraints on the motion of Sgr A* from radio observations \citep{Reid_2020}, giving an apparent motion of $-0.58 \pm 2.23$ km s$^{-1}$ in the direction of Galactic rotation and $-0.85 \pm 0.75$ km s$^{-1}$ toward the North Galactic Pole.
In addition, we verified that the mean square velocity of the SMBH $\langle v^2_{SgrA*} \rangle$ is directly proportional to the mass $m$ of the cluster objects, as expected from energy equipartition between the SMBH and the cluster \citep{Merritt_2007}.


\section{Orbit fitting: deviations from a Schwarzschild orbit}
\label{sec:fitting}
The orbit of S2 around Sgr A* is in remarkable agreement with the predictions of GR, exhibiting a Schwarzschild, in-plane orbital precession of approximately 12 arcmin per orbit. In \cite{GRAVITY_2020, GRAVITY_2022, GRAVITY_2024}, an orbital fit was performed to test the compatibility of the astrometric and spectroscopic data obtained for S2 with a Schwarzschild orbit around Sgr A*, introducing the parameter $f_{SP}$ (see Section \ref{introduction}). The result of \cite{GRAVITY_2024} is $f_{SP}=1.135 \pm 0.110$, where $f_{SP}=0$ corresponds to a Keplerian orbit and $f_{SP}=1$ to a Schwarzschild GR orbit. This indicates that the orbit is compatible with a Schwarzschild orbit at a $\approx 10\sigma$ confidence level.
In addition, an orbital fit was also done assuming $f_{\rm SP}=1$ and a smooth and spherically symmetric extended mass profile, finding an upper limit $M_{e,S2}\,\lesssim\, 1200\,M_\odot$
for the total extended mass enclosed within S2’s orbit \citep{GRAVITY_2024}.

Our goal in this section is to assess whether perturbations caused by a cluster of stellar-mass black holes around Sgr A* could significantly impact the observed orbit of S2, particularly in terms of the measurement of $f_{SP}$ and $M_{e,S2}$. In order to do that, we perform a mock data analysis using the results of the simulations from Section \ref{sec:comp_nbody}, considering the orbit of S2 around Sgr A* in the presence of a cluster of $N$ black holes of equal mass $m$. 
We consider a population of black holes of 20, 50 and 100 M$_\odot$, always such that the total enclosed mass within the apocenter of S2 is $M_{e,S2}=1000 \, \text{M}_\odot$, consistent with the most recent observational constraints \citep{GRAVITY_2024}. 

For each assumed mass of the stellar-mass black holes and each performed simulation, we convert the data into mock observational data. From each simulation, we extract the position $\vec{r}(t)$ and velocity $\vec{v}(t)$ of S2 with respect to Sgr A* in Cartesian coordinates at each time step. For the analysis, we sample approximately 10 data points per year over a full orbital period of S2, starting from apocenter.
The simulated orbit is then projected onto the observer's plane, using the orbital parameters in \cite{GRAVITY_2022}. 
The resulting mock data set consists of the on-sky position of S2, given by its right ascension (RA) and declination (Dec) in arcseconds, and its radial velocity in km s$^{-1}$ as functions of time, mimicking observational data. The star starts its motion at apocenter at $t_{apo} = 2010.35$, consistent with observations.

These data are subsequently used for orbital fitting, where we fit for the six orbital parameters describing the initial osculating Kepler orbit, along with the mass and distance of the central SMBH. Additionally, to test compatibility with GR, we include the parameter $f_{SP}$.
The fitting is performed with a Levenberg-Marquardt $\chi^2$ minimization algorithm, using the same code used for fitting the observational data of S-stars (see \cite{GRAVITY_2018, GRAVITY_2019, GRAVITY_2020} for details).

\begin{figure*}[h!]
        \centering
        \captionsetup[subfigure]{position=top} 
        \begin{subfigure}[b]{0.48\textwidth}
            \centering
            \includegraphics[width=\textwidth]{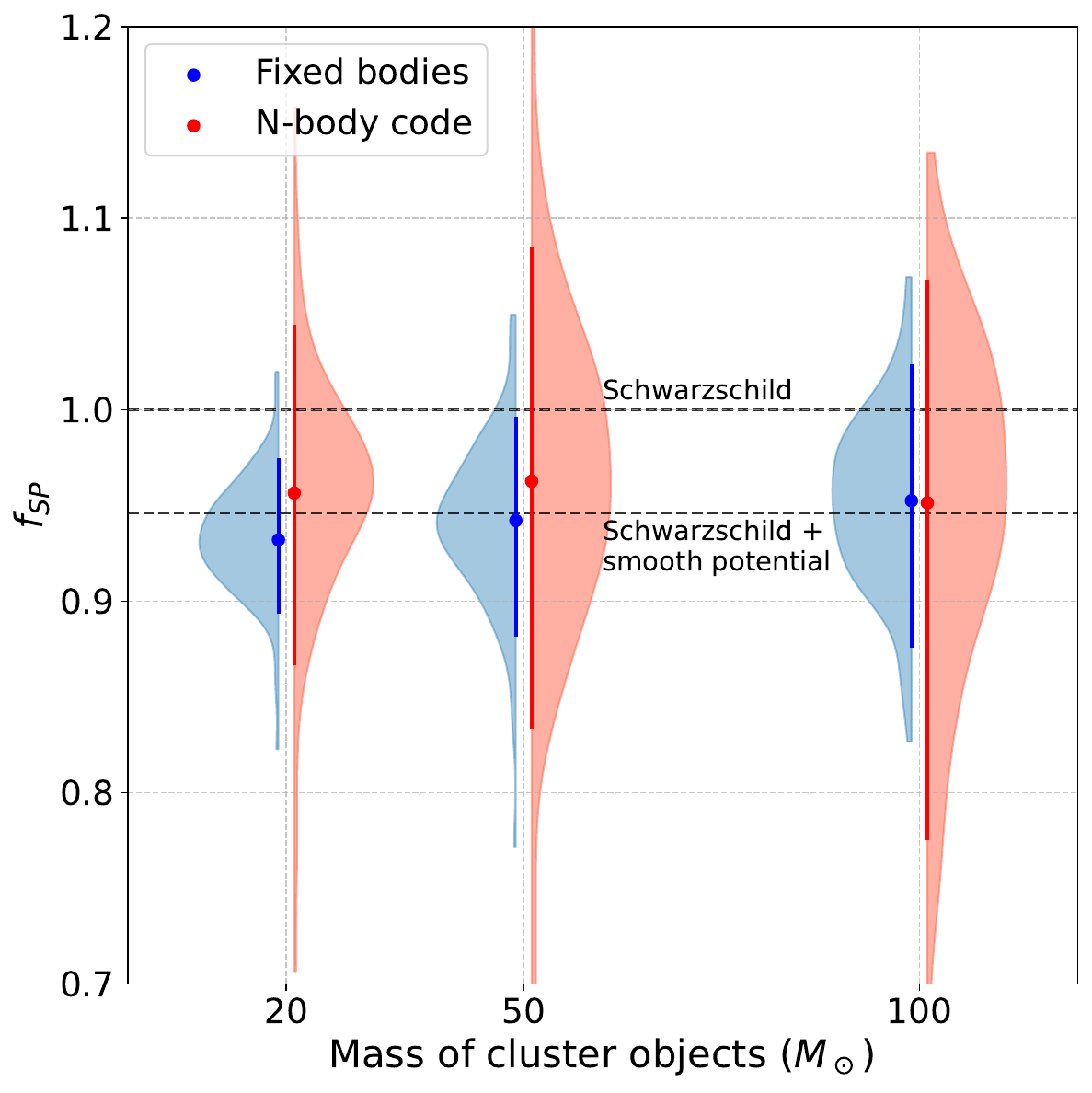}
        \end{subfigure}
        \hfill
        \begin{subfigure}[b]{0.48\textwidth}  
            \centering 
            \includegraphics[width=\textwidth]{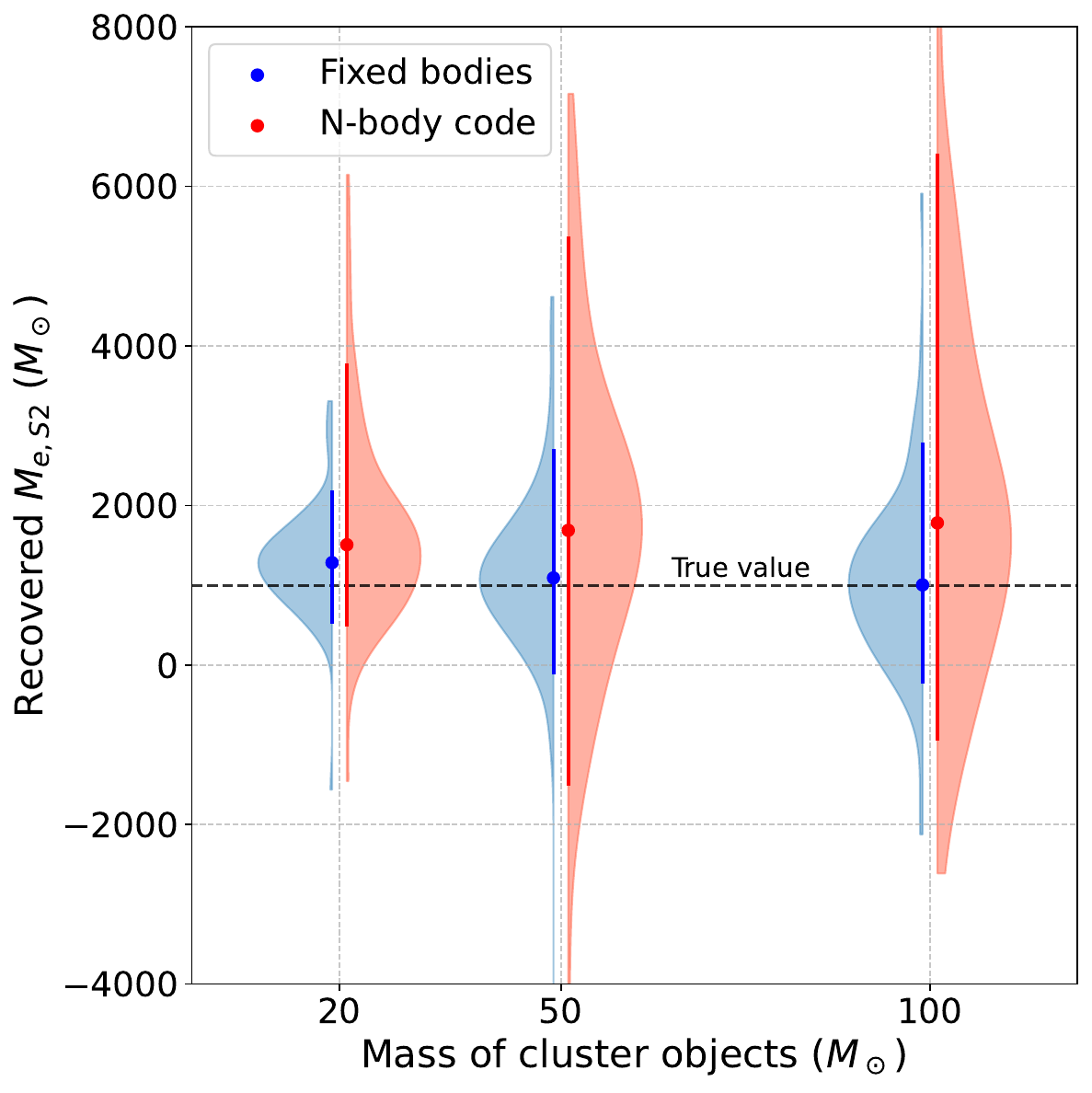}
        \end{subfigure}
        \hfill
        \caption[]
        {Violin plots of the best-fit $f_{\rm sp}$ (left) and $M_{e,S2}$ (right) obtained through a mock data analysis. 
    Results from the simplified approach are shown in blue, while those from the full $N$-body simulations are shown in red. 
        } 
        \label{fig_fsp}%
\end{figure*}

Figure \ref{fig_fsp} (left) displays the results on the distribution of $f_{SP}$, obtained from 100 simulations conducted for each value of the mass of the cluster bodies. We compare again the distributions obtained from the simplified dynamical approach and the full $N$-body code. We observe that the median $f_{SP}$ value is smaller than 1, consistent with the value expected for a GR orbit with a smooth mass distribution of $M_{e,S2}=1000 \, \text{M}_\odot$. However, the range of the $f_{SP}$ distribution is significant, comparable to or exceeding the observational uncertainty of $\approx 0.1$.
For instance, in the case of a distribution of black holes of 20 M$_\odot$, the $f_{SP}$ distribution obtained through $N$-body simulations spans approximately 0.87 to 1.04 (5th to 95th percentile). With fewer but more massive cluster objects the distribution broadens further, ranging approximately between 0.78 and 1.07 in the case of black holes of 100 M$_\odot$. While the distributions obtained from the simplified dynamical approach are slightly narrower than those from $N$-body simulations, the general conclusions remain unchanged. This analysis shows that perturbations from a population of stellar-mass black holes can cause measurable deviations from a Schwarzschild orbit. Accounting for these effects is crucial, particularly given the high precision achievable with the GRAVITY instrument.

In Table \ref{tab:orbital_stats}, we present the fitting results for the specific case of black holes of $m=20 \, \text{M}_\odot$, not only for $f_{SP}$ but also the mass of Sgr A* ($m_\bullet$), its distance ($R_0$), and the six orbital parameters of S2: semi-major axis ($a$), eccentricity ($e$), inclination ($i$), argument of pericenter ($\omega$), longitude of ascending node ($\Omega$), and time of pericenter passage ($t_p$). Our findings show that the mass of Sgr A* can vary by up to approximately 0.13\% from its average value, a variation that is comparable to the observational uncertainty on this parameter. Similar conclusions can be drawn for the distance and the orbital elements of S2. This suggests that the granularity of the mass distribution must be taken into account in order to correctly estimate these parameters.

\begin{table}[h!]
\centering
\caption{Statistics for the fitting parameters in the case of $m = 20 \, \text{M}_\odot$ black holes. For each parameter, we report the mean, standard deviation, range, and the maximum fractional deviation from the mean value ($f_{max}$) across the 100 simulations conducted.}
\label{tab:orbital_stats}
\begin{tabular}{|c|c|c|c|c|}
\hline\hline
Parameter & Mean & Std. dev. & Range & $f_{max}$ (\%) \\
\hline
$f_{sp}$ & 0.955 & 0.0614 & 0.451 & 26\% \\
$m_\bullet$ [$10^6 \text{M}_\odot$]  & 4.30 & 1.70$\times 10^{-3}$ & 9.32$\times 10^{-3}$ & 0.13\% \\
$R_0$ [pc] & 8277 & 1.29 & 5.88 & 0.037\% \\
$a$ [as] & 0.125 & 2.10$\times 10^{-5}$ & 1.17$\times 10^{-4}$ & 0.062\% \\
$e$ & 0.884 & 5.60$\times 10^{-5}$ & 4.42$\times 10^{-4}$ & 0.034\% \\
$i$ [deg] & 134.7 & 5.80$\times 10^{-3}$ & 0.0350 & 0.014\% \\
$\omega$ [deg] & 66.2 & 0.012 & 0.0661 & 0.051\% \\
$\Omega$ [deg] & 228.2 & 1.64$\times 10^{-2}$ & 0.087 & 0.020\% \\
$t_p$ [year] & 2018.37 & 5$\times 10^{-4}$ & 2$\times 10^{-3}$ & $10^{-4}$\% \\
\hline
\end{tabular}
\end{table}

We also fit the mock data under the assumption of $f_{\rm SP}=1$ and a smooth power‐law density distribution,
\begin{equation}
\rho(r)
= \rho_0\!\left(\frac{r}{r_0}\right)^{-2},
\end{equation}
as it is the very profile used to sample the granular potential, estimating the enclosed mass $M_{e,S2}$. The goal is to quantify how a smooth‐potential fit can misestimate $M_{e,S2}$ with respect to the \lq true\rq\ value $M_{e,S2}\approx1000\,M_\odot$, when the distribution is granular. Figure \ref{fig_fsp} (right panel) shows the distribution of the recovered $M_{e,S2}$ for both our simplified dynamical model and the full $N$-body simulations.  As with $f_{\rm SP}$, the distributions obtained from full $N$-body simulations are broader. For cluster objects of $20\,M_\odot$, the 5th–95th percentile range spans $\sim500$ to $3800\,M_\odot$. For $100\,M_\odot$ objects, it expands further, from $\sim-1000$ to $6000\,M_\odot$. 
Thus, granularity in itself can bias the inferred enclosed mass by up to a factor of $\sim6$.  Moreover, for $50$–$100\,M_\odot$ cluster objects a non-negligible fraction of realizations give negative $M_{e,S2}$: in these cases, stochastic close encounters produce a net prograde precession that overwhelms the retrograde precession expected from a smooth mass distribution.
These results demonstrate that fitting the orbit of S2 with a smooth extended potential, when the true background is granular, can lead to wrong or even unphysical mass estimates.  Therefore, future attempts to constrain the extended mass within the orbit of S2 must account for granularity in the stellar‐mass black hole population.

\begin{figure*}[h!]
        \centering
        \captionsetup[subfigure]{position=top} 
                \begin{subfigure}[b]{0.3\textwidth}
            \centering
            \includegraphics[width=\textwidth]{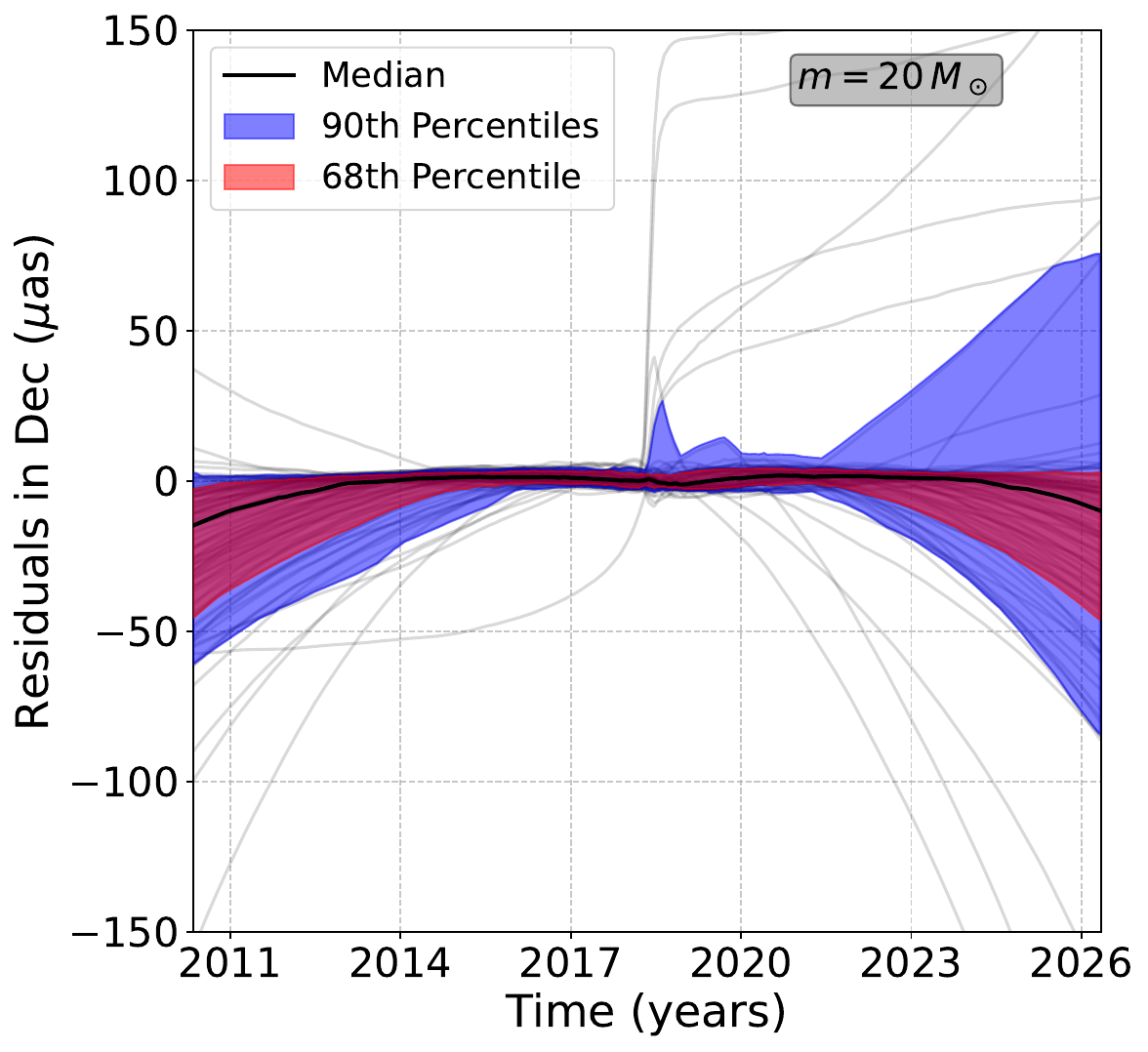}
        \end{subfigure}
        \hfill
        \begin{subfigure}[b]{0.3\textwidth}  
            \centering 
            \includegraphics[width=\textwidth]{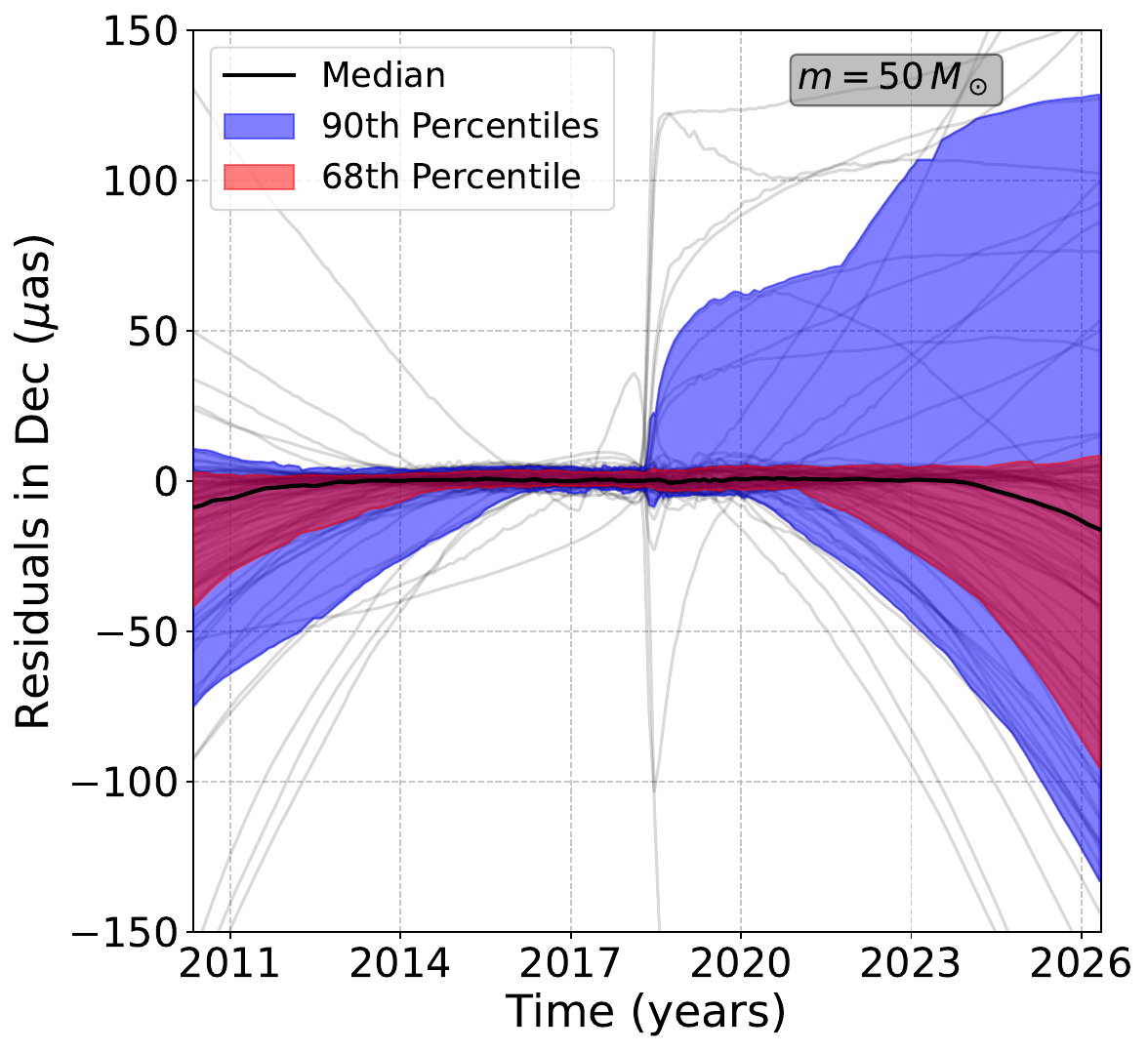}
        \end{subfigure}
        \hfill
        \begin{subfigure}[b]{0.3\textwidth}
            \centering
            \includegraphics[width=\textwidth]{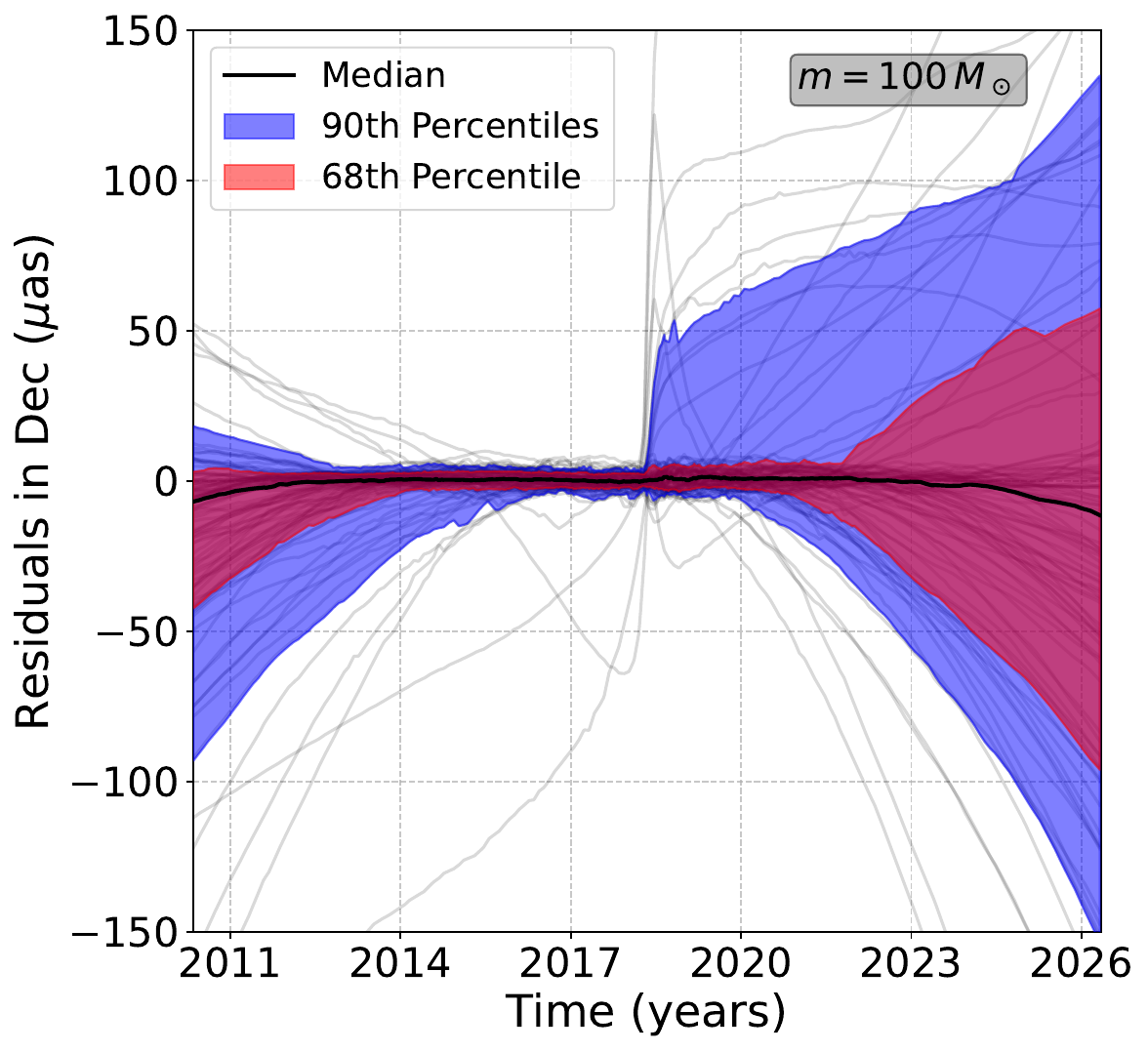}
        \end{subfigure}
        \hfill
        \begin{subfigure}[b]{0.3\textwidth}
            \centering
            \includegraphics[width=\textwidth]{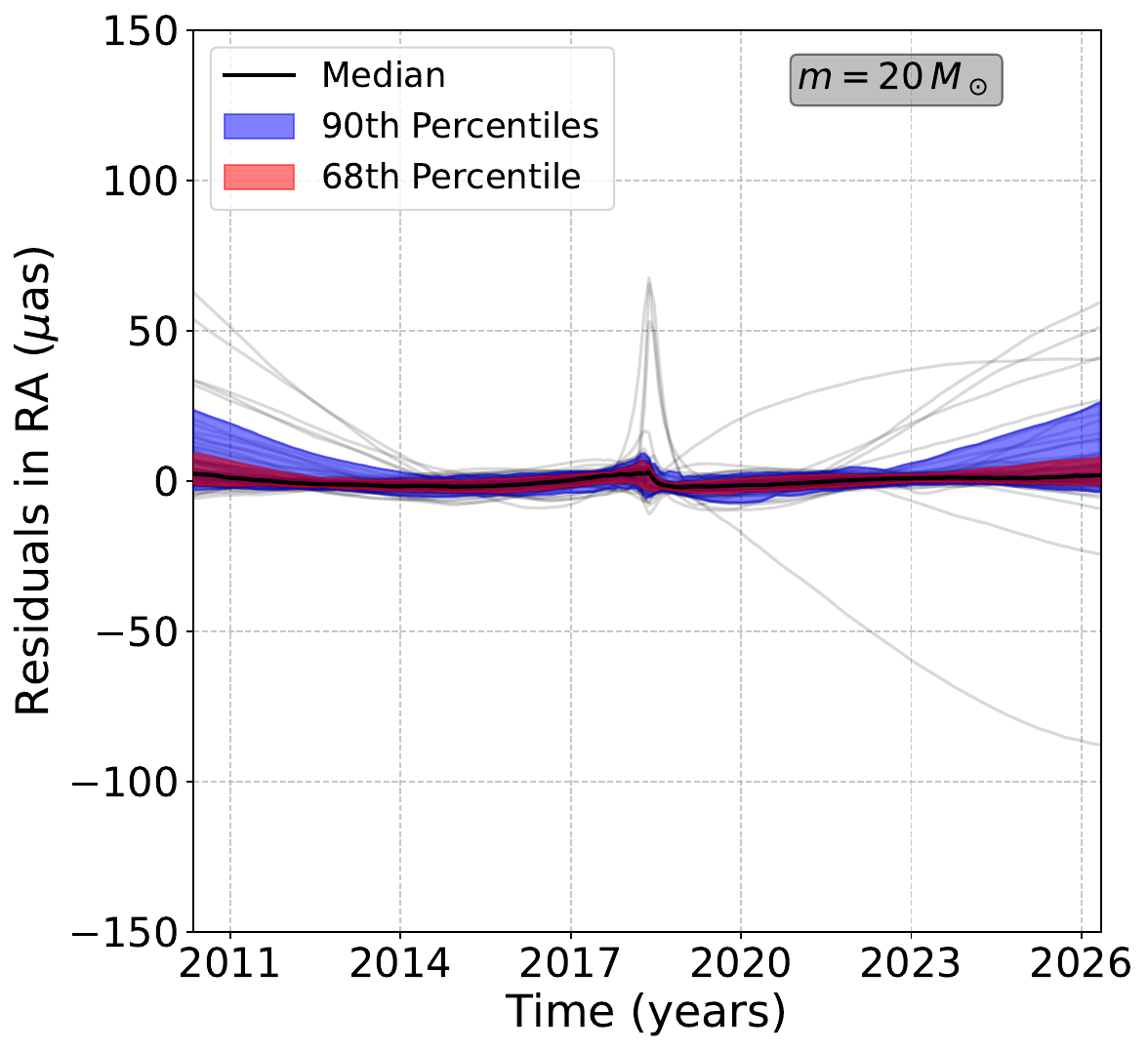}
        \end{subfigure}
        \hfill
        \begin{subfigure}[b]{0.3\textwidth}
            \centering
            \includegraphics[width=\textwidth]{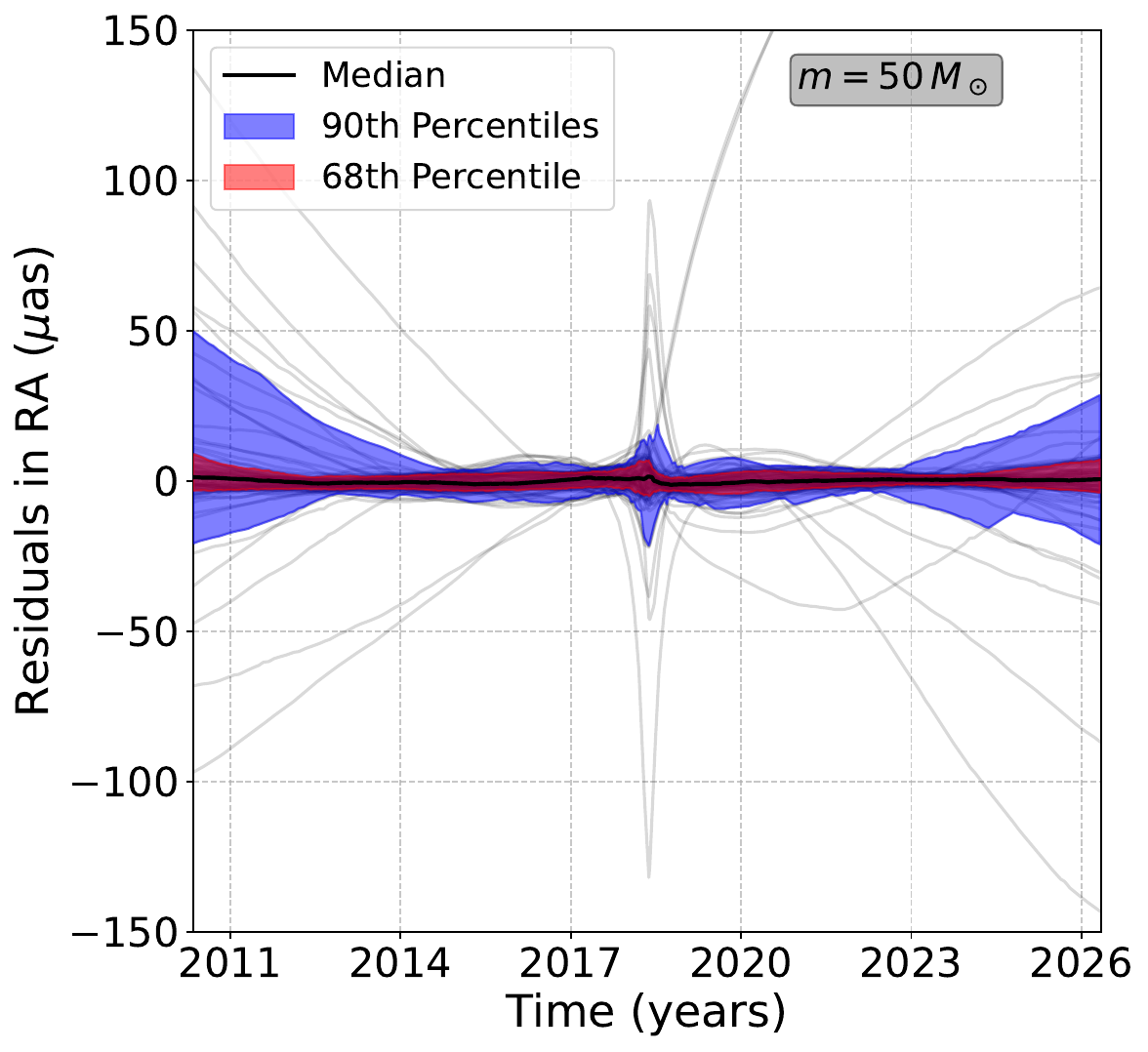}
        \end{subfigure}
        \hfill
        \begin{subfigure}[b]{0.3\textwidth}
            \centering
            \includegraphics[width=\textwidth]{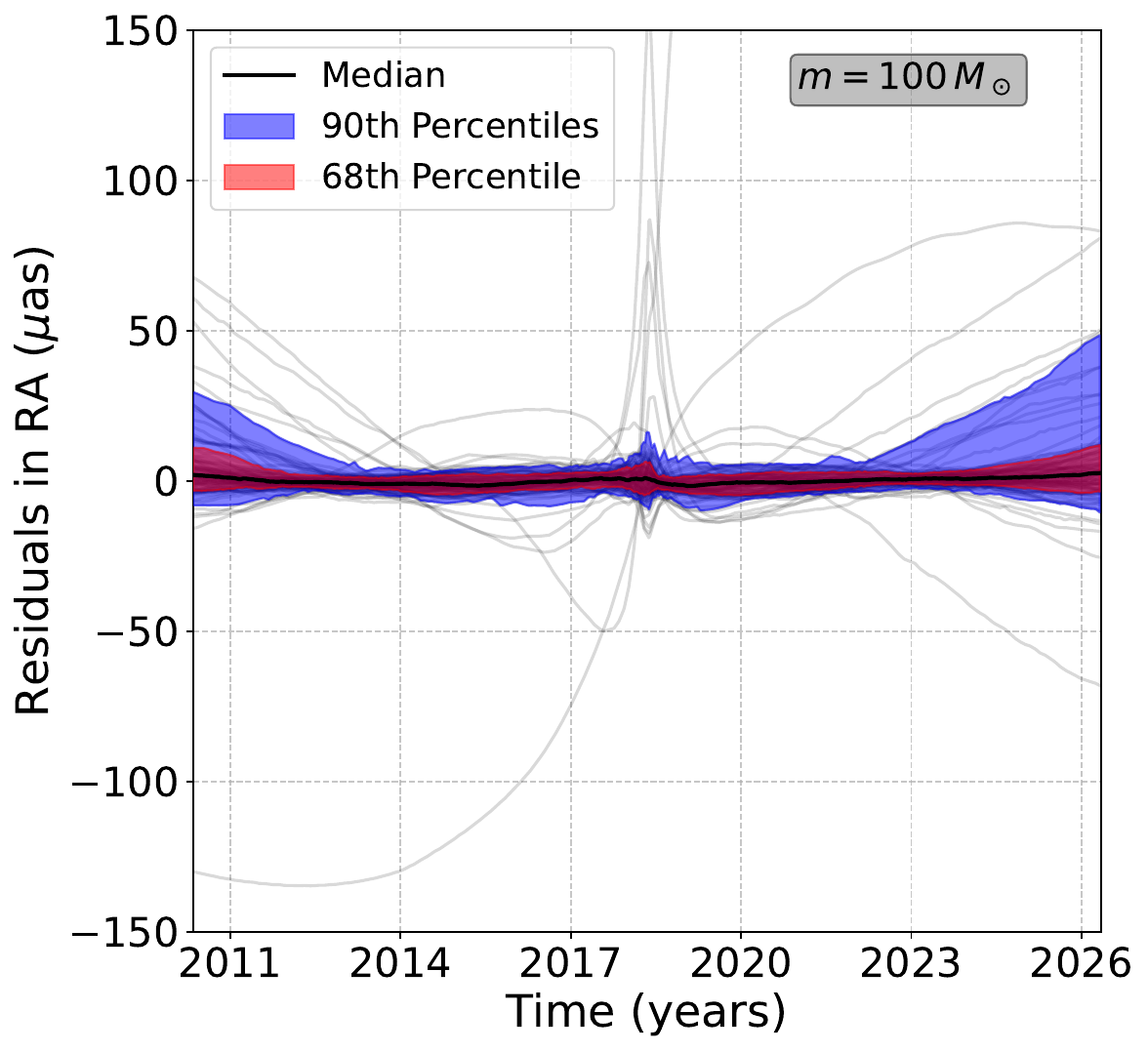}
        \end{subfigure}
        \hfill
                \begin{subfigure}[b]{0.3\textwidth}
            \centering
            \includegraphics[width=\textwidth]{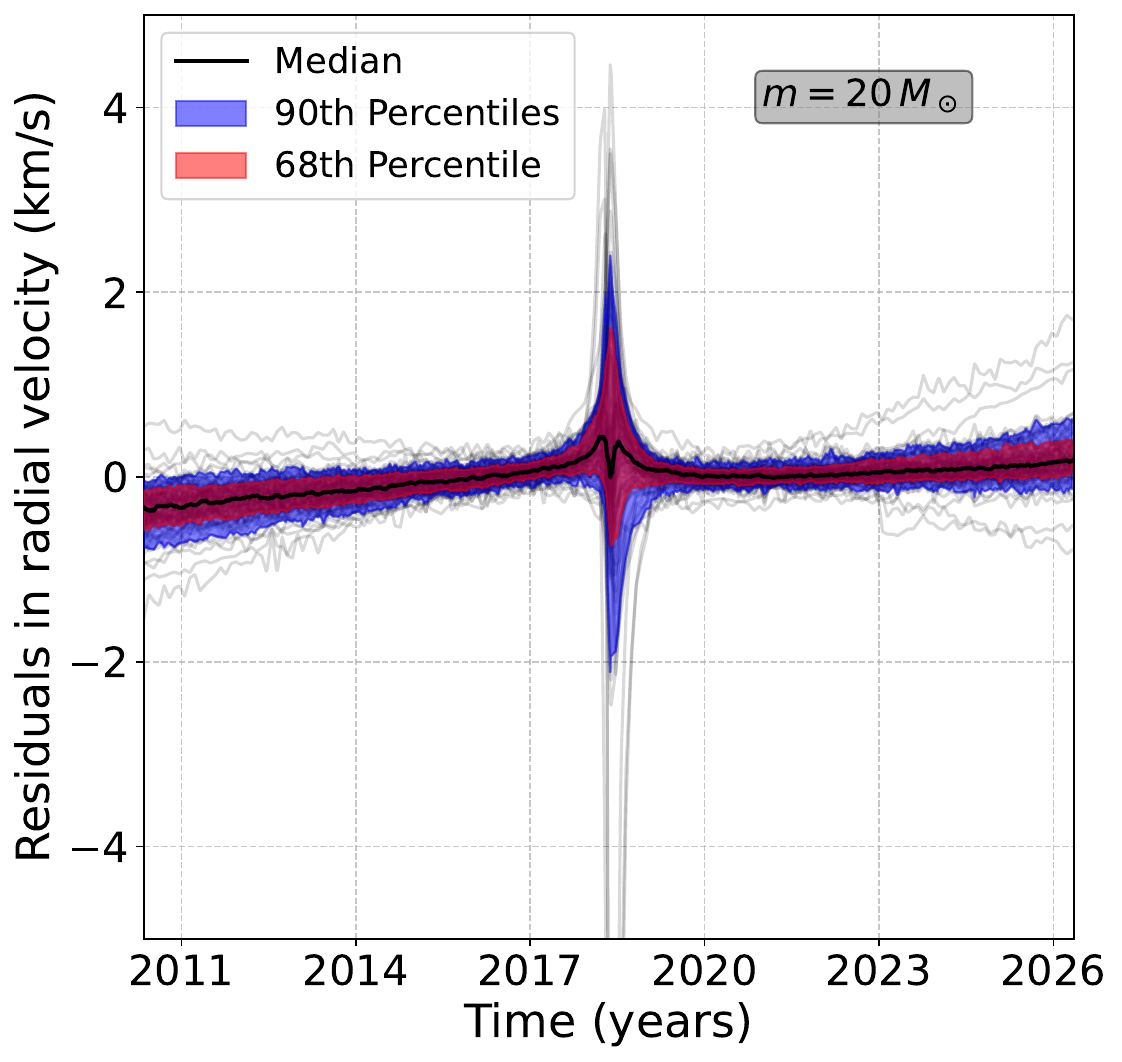}
        \end{subfigure}
        \hfill
        \begin{subfigure}[b]{0.3\textwidth}  
            \centering 
            \includegraphics[width=\textwidth]{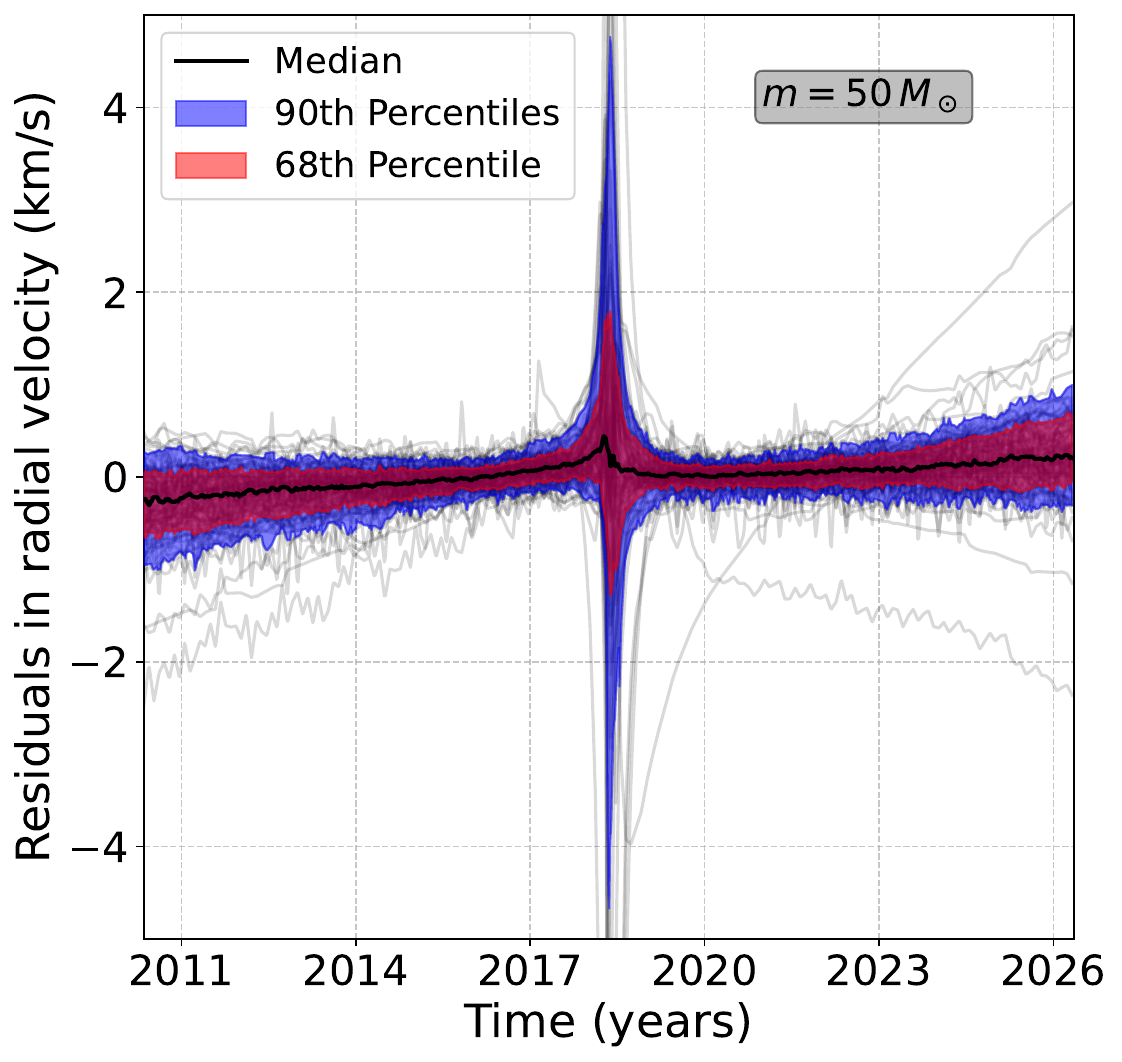}
        \end{subfigure}
        \hfill
        \begin{subfigure}[b]{0.3\textwidth}
            \centering
            \includegraphics[width=\textwidth]{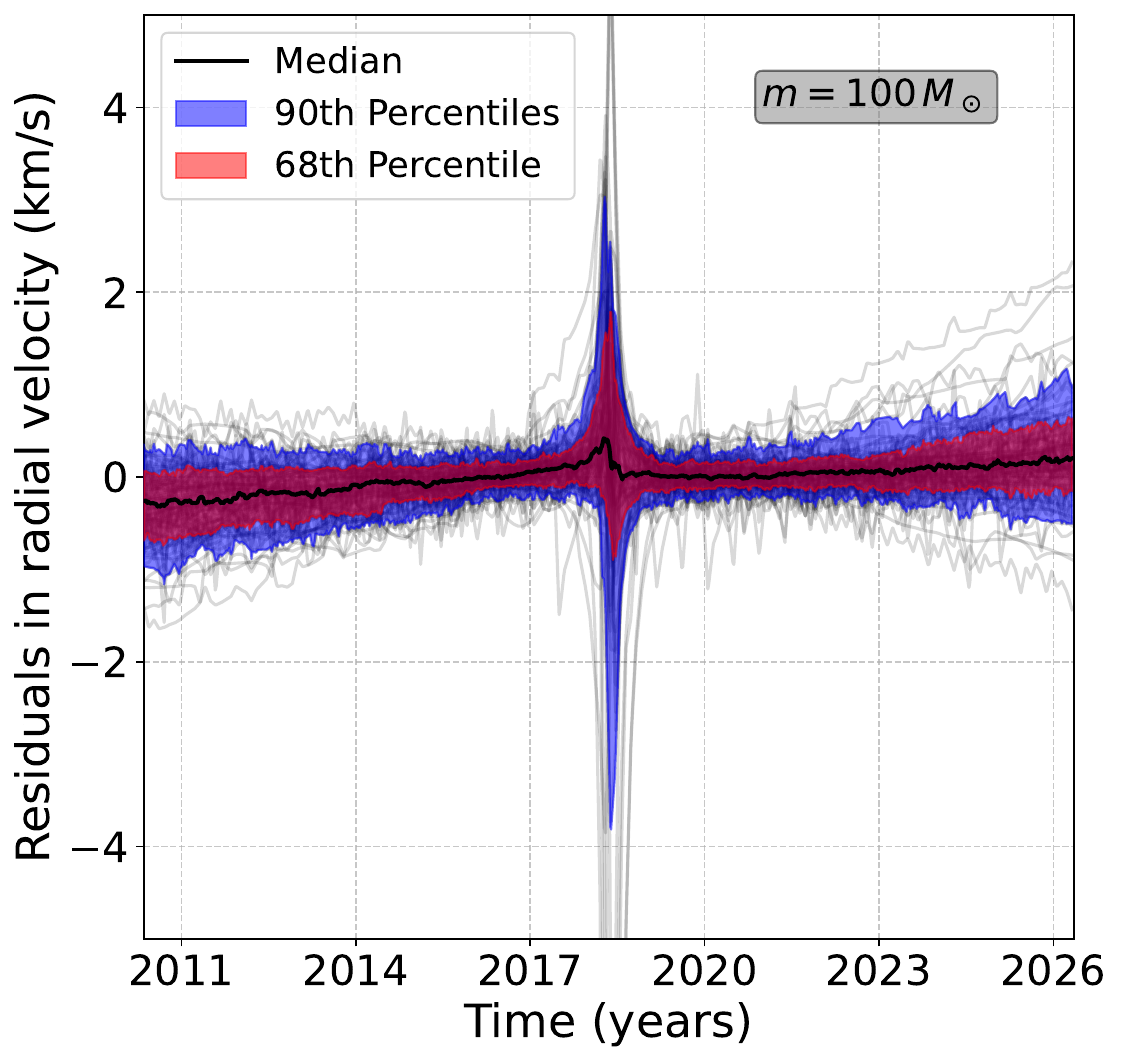}
        \end{subfigure}
        \hfill
        \caption[]
        {Residuals in Dec (first row), RA (second row) and radial velocity (third row) as functions of time, between the 100 simulated orbits in the granular case (full $N$-body simulations) and the respective best-fit Schwarzschild orbits ($f_{SP}=1$). The first column gives the results considering a population of $20 \, \text{M}_\odot$ black holes, the second column $50 \, \text{M}_\odot$ black holes, the third $100 \, \text{M}_\odot$ black holes. Each of the 100 residual curves is shown in light grey, highlighting their median (solid black line), 68th percentile (red filled area) and 90th percentile (blue filled area).
        }
        \label{fig:astr_residuals}
        \end{figure*}

In addition, we also analyze the residuals in astrometry and radial velocity between each simulated orbit and the respective best-fit Schwarzschild orbit (with $f_{SP}$ fixed to 1). 
In Fig. \ref{fig:astr_residuals} we plot the residuals in Dec, RA and radial velocity as a function of time for the 100 full $N$-body simulations conducted, for each of the three different black hole masses considered. In each plot, we highlight the median, the 68th percentile and 90th percentile of the distribution.
We stress that, to isolate the true observational signature of the perturbations by a cluster of stellar‐mass black holes, it is essential to fit each mock data set with a relativistic orbit and then examine the residuals.  Simply taking the instantaneous difference between a perturbed trajectory and the unperturbed one would introduce artificial peaks at pericenter: because perturbations slightly change the orbital period, the two trajectories dephase and produce large spatial offsets where the orbital velocity is highest.  By contrast, real observations do not sample an instantaneous offset between two fixed trajectories, but rather measure a single set of astrometric and spectroscopic data that must be fitted to a model orbit.  By fitting the mock data, any phase drift is absorbed and the residuals accurately reflect where perturbations are actually measurable.

We find that the astrometric residuals are strongly time-dependent, with the largest astrometric deviations occurring almost always near apocenter.  This can be naturally explained by the fact that the star spends more time in the apocenter half of its orbit, where the gravitational force exerted by the SMBH is weaker, allowing perturbations to play a more significant role. Indeed, for $20$ M$_\odot$ perturbers the average force ratio $\left\langle f_{\rm gran}/f_{\rm MBH}\right\rangle$ at apocenter is $\sim15$ times larger than at pericenter. In addition, at apocenter the star is at larger separation from Sgr A*, and so deviations are `magnified'. 

In the apocenter half of the orbit, the residuals are significantly larger in Dec than in RA. This can be explained by the geometry of the orbit of S2, as the star explores a broader range in Dec.
Looking at the astrometric residuals in 2022.7, which is the time corresponding to the most recent data of S2 analyzed in \citet{GRAVITY_2024}, we notice that they are relatively small, with a 68\% probability of being smaller than 30 \text{$\mu as$}, which is around the astrometric accuracy of GRAVITY.
This may explain why the orbit of S2 in \citet{GRAVITY_2024} is in perfect agreement with a Schwarzschild orbit, as it is very likely that deviations caused by the granular mass distribution are below the accuracy of GRAVITY. 
However, looking at the residuals at the time of the next apocenter passage of S2 in 2026.35, the residuals in Dec can be significantly larger than the astrometric accuracy of GRAVITY. They are larger than 30 \text{$\mu as$} in 35\% (for black holes of 20 M$_\odot$) to 60\% (for 100 M$_\odot$) of the simulations. In around 10\% (for 20 M$_\odot$) to 25\% (for 100 M$_\odot$) of the simulations performed the residuals are even larger than 100 \text{$\mu as$}. 
These results give hope of detecting a deviation of the S2 orbit from a pure Schwarzschild orbit predicted by GR with future astrometric observations. In fact, observing a residual in Dec around the time of the next apocenter passage of S2 in 2026.35, the first to be observed with GRAVITY, could provide direct evidence of scattering effects on the orbit of S2 induced by a population of stellar-mass black holes. Actually, by 2026 the upgrade of GRAVITY to GRAVITY+ at the VLTI \citep{GRAVITY+} will be completed, potentially reaching an even higher astrometric accuracy.

Finally, residuals in radial velocity are largest at pericenter. However, in almost all cases they are below the precision of current instrumentation, such as the ERIS spectrograph at the VLT, which achieves an accuracy of approximately 7 km s$^{-1}$ for S2. 

\section{Conclusions}
\label{concl}
In this paper we analyzed the impact of the granularity of the mass distribution, supposed as a cluster of equal-mass objects, on the orbit of S2 around Sagittarius A*, using a statistically reliable set of simulations.

The main conclusions of this study are:

\begin{itemize}
    \item[$\bullet$] The presence of a granular mass distribution leads to deviations from the motion in case of a smooth potential, 
    primarily driven by local scattering events,  causing precession of the orbital osculatory plane and a variation of the in-plane angular precession. 
    \item[$\bullet$] Larger deviations occur when considering larger total mass of the granular distribution, as well as a larger mass of the individual cluster objects. 
    For an enclosed mass of $100 \, \text{M}_\odot$ within the apocenter of S2, the average precession of the orbital plane reaches up to 0.2 arcmin. For an enclosed mass of $1000 \, \text{M}_\odot$, it reaches up to 0.2 arcmin for $1 \, \text{M}_\odot$ cluster objects and up to 1.5 arcmin for $100 \, \text{M}_\odot$ objects. In the latter case, the in-plane precession can deviate by up to 1.5 arcmin, corresponding to a fractional variation of 13\%.
    \item[$\bullet$] Our results, obtained with a simplified dynamical approach, have been validated, for a subset of cases, with full $N$-body simulations. While the in-plane precession results obtained in the two ways are fully consistent, the $N$-body simulations reveal larger orbital plane precession. This is attributed to the ``Brownian motion'' of the SMBH, induced by the interaction with the cluster objects, which is absent in the simplified approach. The SMBH experiences a mean displacement up to 6 \text{$\mu as$} and a mean velocity up to 238 m s$^{-1}$ over one radial period of S2, in the case of 100 M$_\odot$ cluster objects.
    \item[$\bullet$] Mock data analysis reveals that perturbations caused by stellar-mass black holes on the orbit of S2 could produce observable deviations from a Schwarzschild orbit. The parameter $f_{SP}$, which quantifies deviations from a Schwarzschild orbit, can deviate significantly from 1 in the presence of a granular distribution, exceeding the current observational uncertainty of $\approx 0.1$.
    The astrometric residuals between the simulated orbits and the respective best-fit Schwarzschild orbits are strongly time-dependent. 
    Around apocenter, the residuals in Declination exceed the accuracy of GRAVITY in more than 35 to 60 \% of simulations, assuming black holes with masses between $20$ and $100 \,$ M$_\odot$. This makes the upcoming apocenter passage of S2 in 2026 a unique opportunity to detect, for the first time, scattering effects on the orbit of S2 caused by stellar-mass black holes. 
    Residuals in radial velocity, instead, are largest at pericenter, but are typically below the precision achievable with current instrumentation.
    \item[$\bullet$] Mock data analysis further reveals that fitting the orbit of S2 under the assumption of $f_{SP}=1$ (i.e. Schwarzschild orbit) and a smooth extended mass profile (as done in \citealp{GRAVITY_2024}), when the true distribution is granular, can result in wrong or even unphysical mass estimates.  The inferred enclosed mass within S2’s orbit, $M_{e,S2}$, can deviate by up to a factor of $\sim6$ from the true value and may even be negative.  Thus, any attempt to constrain the extended mass within the orbit of S2 must explicitly account for granularity in the stellar‐mass black hole population.
\end{itemize}

As a final remark, the detection of perturbations caused by stellar-mass black holes on the orbit of S2 would have profound implications beyond the immediate study of GC dynamics. Actually, such observations would provide direct evidence for the presence of stellar-mass black holes in the GC region, which are predicted to be progenitors of EMRIs \citep{LISA_2017}. While the merger rate of EMRIs in the GC is expected to be negligible, a significant population of stellar black holes in their early inspiral phase, known as early EMRIs, could still be detected by LISA. These detections would enable measurements of the mass and spin of Sgr A* with remarkable accuracy \citep{Pau_2024, AmaroSeoane2025a, AmaroSeoane2025b}. 
Future observations with GRAVITY and its upgraded version GRAVITY+, which will be operational in 2026, will probe the hidden population of stellar‐mass black holes years before LISA’s launch, helping to determine whether early EMRIs in the GC will be detectable and providing an observational constraint on the EMRI rate in Milky Way–like galaxies.
In addition, before the launch of LISA, the MICADO instrument at the ELT \citep{MICADO_ELT} will enable the observation of much fainter stars than currently possible with GRAVITY, potentially identifying stars in tighter orbits around Sgr A*. Such stars could provide an opportunity to measure the spin of Sgr A* \citep{Merritt_2010, Waisb_2018, Cap_Sad_2023}. However, as noted in \citet{Merritt_2010}, perturbations by stellar-mass black holes may complicate the detection of the spin. Nonetheless, the Lense-Thirring effect is strongly peaked at pericenter, offering hope of disentangling the influence of the spin from the effects of stellar perturbations \citep{Zhang_Iorio_2017}.
  

\begin{acknowledgements}
We thank Pau Amaro-Seoane for the helpful comments that improved this paper.
AG was supported at the Technion by a Zuckerman Fellowship. This work made use of the following software packages: Astropy \citep{astropy+2013, astropy+2018, astropy+2022}, ARWV \citep{chacd19}, Jupyter \citep{kluyver+2016}, Matplotlib \citep{hunter2007}, NumPy \citep{harris+2020Numpy}, pandas \citep{pandas2010}, REBOUND \citep{rein.liu2012}, SciPy \citep{virtanen+2020}.
\end{acknowledgements}

\balance
\bibliography{aa_submission}

\begin{thebibliography}{54}
\expandafter\ifx\csname natexlab\endcsname\relax\def\natexlab#1{#1}\fi

\bibitem[{Abbott {et~al.}(2023)Abbott, Abbott, Acernese, Ackley, Adams,
  Adhikari, Adhikari, Adya, Affeldt, Agarwal, Agathos, Agatsuma, Aggarwal,
  Aguiar, Aiello, Ain, Ajith, Akutsu, de~Alarc\'on, Akcay, Albanesi, Allocca,
  Altin, Amato, Anand, Anand, Ananyeva, Anderson, Anderson, Ando, Andrade,
  Andres, Andri\ifmmode~\acute{c}\else \'{c}\fi{}, Angelova, Ansoldi, Antelis,
  Antier, Antonini, Appert, Arai, Arai, Arai, Araki, Araya, Araya, Areeda,
  Ar\`ene, Aritomi, Arnaud, Arogeti, Aronson, Arun, Asada, Asali, Ashton, Aso,
  Assiduo, Aston, Astone, Aubin, Austin, Babak, Badaracco, Bader, Badger, Bae,
  Bae, Baer, Bagnasco, Bai, Baiotti, Baird, Bajpai, Ball, Ballardin, Ballmer,
  Balsamo, Baltus, Banagiri, Bankar, Barayoga, Barbieri, Barish, Barker,
  Barneo, Barone, Barr, Barsotti, Barsuglia, Barta, Bartlett, Barton, Bartos,
  Bassiri, Basti, Bawaj, Bayley, Baylor, Bazzan, B\'ecsy, Bedakihale, Bejger,
  Belahcene, Benedetto, Beniwal, Bennett, Bentley, BenYaala, Bergamin, Berger,
  Bernuzzi, Berry, Bersanetti, Bertolini, Betzwieser, Beveridge, Bhandare,
  Bhardwaj, Bhattacharjee, Bhaumik, Bilenko, Billingsley, Bini, Birney,
  Birnholtz, Biscans, Bischi, Biscoveanu, Bisht, Biswas, Bitossi, Bizouard,
  Blackburn, Blair, Blair, Blair, Bobba, Bode, Boer, Bogaert, Boldrini,
  Bonavena, Bondu, Bonilla, Bonnand, Booker, Boom, Bork, Boschi, Bose, Bose,
  Bossilkov, Boudart, Bouffanais, Bozzi, Bradaschia, Brady, Bramley, Branch,
  Branchesi, Brandt, Brau, Breschi, Briant, Briggs, Brillet, Brinkmann,
  Brockill, Brooks, Brooks, Brown, Brunett, Bruno, Bruntz, Bryant, Bulik,
  Bulten, Buonanno, Buscicchio, Buskulic, Buy, Byer, Cadonati, Cagnoli,
  Cahillane, Bustillo, Callaghan, Callister, Calloni, Cameron, Camp, Canepa,
  Canevarolo, Cannavacciuolo, Cannon, Cao, Cao, Capocasa, Capote, Carapella,
  Carbognani, Carlin, Carney, Carpinelli, Carrillo, Carullo, Carver, Diaz,
  Casentini, Castaldi, Caudill, Cavagli\`a, Cavalier, Cavalieri, Ceasar, Cella,
  Cerd\'a-Dur\'an, Cesarini, Chaibi, Chakravarti, Subrahmanya, Champion, Chan,
  Chan, Chan, Chan, Chan, Chandra, Chanial, Chao, Chapman-Bird, Charlton,
  Chase, Chassande-Mottin, Chatterjee, Chatterjee, Chatterjee, Chaturvedi,
  Chaty, Chatziioannou, Chen, Chen, Chen, Chen, Chen, Chen, Chen, Chen, Cheng,
  Cheong, Cheung, Chia, Chiadini, Chiang, Chiarini, Chierici, Chincarini,
  Chiofalo, Chiummo, Cho, Cho, Choudhary, Choudhary, Christensen, Chu, Chu,
  Chu, Chua, Chung, Ciani, Ciecielag, Cie\ifmmode~\acute{s}\else \'{s}\fi{}lar,
  Cifaldi, Ciobanu, Ciolfi, Cipriano, Cirone, Clara, Clark, Clark, Clarke,
  Clearwater, Clesse, Cleva, Coccia, Codazzo, Cohadon, Cohen, Cohen, Colleoni,
  Collette, Colombo, Colpi, Compton, Constancio, Conti, Cooper, Corban,
  Corbitt, Cordero-Carri\'on, Corezzi, Corley, Cornish, Corre, Corsi, Cortese,
  Costa, Cotesta, Coughlin, Coulon, Countryman, Cousins, Couvares, Coward,
  Cowart, Coyne, Coyne, Creighton, Creighton, Criswell, Croquette, Crowder,
  Cudell, Cullen, Cumming, Cummings, Cunningham, Cuoco, Cury\l{}o, Dabadie,
  Canton, Dall'Osso, D\'alya, Dana, DaneshgaranBajastani, D'Angelo, Danila,
  Danilishin, D'Antonio, Danzmann, Darsow-Fromm, Dasgupta, Datrier, Datta,
  Dattilo, Dave, Davier, Davies, Davis, Davis, Daw, Dean, DeBra, Deenadayalan,
  Degallaix, De~Laurentis, Del\'eglise, Del~Favero, De~Lillo, De~Lillo,
  Del~Pozzo, DeMarchi, De~Matteis, D'Emilio, Demos, Dent, Depasse, De~Pietri,
  De~Rosa, De~Rossi, DeSalvo, De~Simone, Dhurandhar, D\'{\i}az, Diaz-Ortiz,
  Didio, Dietrich, Di~Fiore, Di~Fronzo, Di~Giorgio, Di~Giovanni, Di~Giovanni,
  Di~Girolamo, Di~Lieto, Ding, Di~Pace, Di~Palma, Di~Renzo, Divakarla,
  Dmitriev, Doctor, D'Onofrio, Donovan, Dooley, Doravari, Dorrington, Drago,
  Driggers, Drori, Ducoin, Dupej, Durante, D'Urso, Duverne, Dwyer, Eassa,
  Easter, Ebersold, Eckhardt, Eddolls, Edelman, Edo, Edy, Effler, Eguchi,
  Eichholz, Eikenberry, Eisenmann, Eisenstein, Ejlli, Engelby, Enomoto, Errico,
  Essick, Estell\'es, Estevez, Etienne, Etzel, Evans, Evans, Ewing, Fafone,
  Fair, Fairhurst, Farah, Farinon, Farr, Farr, Farrow, Fauchon-Jones, Favaro,
  Favata, Fays, Fazio, Feicht, Fejer, Fenyvesi, Ferguson, Fernandez-Galiana,
  Ferrante, Ferreira, Fidecaro, Figura, Fiori, Fishbach, Fisher, Fittipaldi,
  Fiumara, Flaminio, Floden, Fong, Font, Fornal, Forsyth, Franke, Frasca,
  Frasconi, Frederick, Freed, Frei, Freise, Frey, Fritschel, Frolov, Fronz\'e,
  Fujii, Fujikawa, Fukunaga, Fukushima, Fulda, Fyffe, Gabbard, Gadre, Gair,
  Gais, Galaudage, Gamba, Ganapathy, Ganguly, Gao, Gaonkar, Garaventa,
  Garc\'{\i}a, Garc\'{\i}a-N\'u\~nez, Garc\'{\i}a-Quir\'os, Garufi, Gateley,
  Gaudio, Gayathri, Ge, Gemme, Gennai, George, George, Gerberding, Gergely,
  Gewecke, Ghonge, Ghosh, Ghosh, Ghosh, Ghosh, Giacomazzo, Giacoppo, Giaime,
  Giardina, Gibson, Gier, Giesler, Giri, Gissi, Glanzer, Gleckl, Godwin,
  Golomb, Goetz, Goetz, Gohlke, Goncharov, Gonz\'alez, Gopakumar, Gosselin,
  Gouaty, Gould, Grace, Grado, Granata, Granata, Grant, Gras, Grassia, Gray,
  Gray, Greco, Green, Green, Gretarsson, Gretarsson, Griffith, Griffiths,
  Griggs, Grignani, Grimaldi, Grimm, Grote, Grunewald, Gruning, Guerra, Guidi,
  Guimaraes, Guix\'e, Gulati, Guo, Guo, Gupta, Gupta, Gupta, Gustafson,
  Gustafson, Guzman, Ha, Haegel, Hagiwara, Haino, Halim, Hall, Hamilton,
  Hammond, Han, Haney, Hanks, Hanna, Hannam, Hannuksela, Hansen, Hansen,
  Hanson, Harder, Hardwick, Haris, Harms, Harry, Harry, Hartwig, Hasegawa,
  Haskell, Hasskew, Haster, Hattori, Haughian, Hayakawa, Hayama, Hayes, Healy,
  Heidmann, Heidt, Heintze, Heinze, Heinzel, Heitmann, Hellman, Hello,
  Helmling-Cornell, Hemming, Hendry, Heng, Hennes, Hennig, Hennig, Hernandez,
  Vivanco, Heurs, Hild, Hill, Himemoto, Hines, Hiranuma, Hirata, Hirose,
  Hochheim, Hofman, Hohmann, Holcomb, Holland, Hollows, Holmes, Holt, Holz,
  Hong, Hopkins, Hough, Hourihane, Howell, Hoy, Hoyland, Hreibi, Hsieh, Hsu,
  Huang, Huang, Huang, Huang, Huang, Huang, H\"ubner, Huddart, Hughey, Hui,
  Hui, Husa, Huttner, Huxford, Huynh-Dinh, Ide, Idzkowski, Iess, Ikenoue, Imam,
  Inayoshi, Ingram, Inoue, Ioka, Isi, Isleif, Ito, Itoh, Iyer, Izumi,
  JaberianHamedan, Jacqmin, Jadhav, Jadhav, James, Jan, Jani, Janquart,
  Janssens, Janthalur, Jaranowski, Jariwala, Jaume, Jenkins, Jenner, Jeon,
  Jeunon, Jia, Jin, Johns, Jones, Jones, Jones, Jones, Jones, Jonker, Ju, Jung,
  Jung, Junker, Juste, Kaihotsu, Kajita, Kakizaki, Kalaghatgi, Kalogera, Kamai,
  Kamiizumi, Kanda, Kandhasamy, Kang, Kanner, Kao, Kapadia, Kapasi, Karat,
  Karathanasis, Karki, Kashyap, Kasprzack, Kastaun, Katsanevas, Katsavounidis,
  Katzman, Kaur, Kawabe, Kawaguchi, Kawai, Kawasaki, K\'ef\'elian, Keitel, Key,
  Khadka, Khalili, Khan, Khazanov, Khetan, Khursheed, Kijbunchoo, Kim, Kim,
  Kim, Kim, Kim, Kim, Kimball, Kimura, Kinley-Hanlon, Kirchhoff, Kissel, Kita,
  Kitazawa, Kleybolte, Klimenko, Knee, Knowles, Knyazev, Koch, Koekoek, Kojima,
  Kokeyama, Koley, Kolitsidou, Kolstein, Komori, Kondrashov, Kong, Kontos,
  Koper, Korobko, Kotake, Kovalam, Kozak, Kozakai, Kozu, Kringel, Krishnendu,
  Kr\'olak, Kuehn, Kuei, Kuijer, Kulkarni, Kumar, Kumar, Kumar, Kumar, Kume,
  Kuns, Kuo, Kuo, Kuromiya, Kuroyanagi, Kusayanagi, Kuwahara, Kwak, Lagabbe,
  Laghi, Lalande, Lam, Lamberts, Landry, Landry, Lane, Lang, Lange, Lantz,
  La~Rosa, Lartaux-Vollard, Lasky, Laxen, Lazzarini, Lazzaro, Leaci, Leavey,
  Lecoeuche, Lee, Lee, Lee, Lee, Lee, Lee, Lehmann, Lema\^{\i}tre, Leonardi,
  Leroy, Letendre, Levesque, Levin, Leviton, Leyde, Li, Li, Li, Li, Li, Li,
  Lin, Lin, Lin, Lin, Lin, Linde, Linker, Linley, Littenberg, Liu, Liu, Liu,
  Liu, Llamas, Llorens-Monteagudo, Lo, Lockwood, Loh, London, Longo, Lopez,
  Portilla, Lorenzini, Loriette, Lormand, Losurdo, Lott, Lough, Lousto,
  Lovelace, Lucaccioni, L\"uck, Lumaca, Lundgren, Luo, Lynam, Macas, MacInnis,
  Macleod, MacMillan, Macquet, Hernandez, Magazz\`u, Magee, Maggiore, Magnozzi,
  Mahesh, Majorana, Makarem, Maksimovic, Maliakal, Malik, Man, Mandic, Mangano,
  Mango, Mansell, Manske, Mantovani, Mapelli, Marchesoni, Marchio, Marion,
  Mark, M\'arka, M\'arka, Markakis, Markosyan, Markowitz, Maros, Marquina,
  Marsat, Martelli, Martin, Martin, Martinez, Martinez, Martinez, Martinovic,
  Martynov, Marx, Masalehdan, Mason, Massera, Masserot, Massinger, Masso-Reid,
  Mastrogiovanni, Matas, Mateu-Lucena, Matichard, Matiushechkina, Mavalvala,
  McCann, McCarthy, McClelland, McClincy, McCormick, McCuller, McGhee, McGuire,
  McIsaac, McIver, McRae, McWilliams, Meacher, Mehmet, Mehta, Meijer, Melatos,
  Melchor, Mendell, Menendez-Vazquez, Menoni, Mercer, Mereni, Merfeld, Merilh,
  Merritt, Merzougui, Meshkov, Messenger, Messick, Meyers, Meylahn, Mhaske,
  Miani, Miao, Michaloliakos, Michel, Michimura, Middleton, Milano, Miller,
  Miller, Miller, Miller, Millhouse, Mills, Milotti, Minazzoli, Minenkov, Mio,
  Mir, Miravet-Ten\'es, Mishra, Mishra, Mistry, Mitra, Mitrofanov,
  Mitselmakher, Mittleman, Miyakawa, Miyamoto, Miyazaki, Miyo, Miyoki, Mo,
  Modafferi, Moguel, Mogushi, Mohapatra, Mohite, Molina, Molina-Ruiz, Mondin,
  Montani, Moore, Moraru, Morawski, More, Moreno, Moreno, Mori, Morisaki,
  Moriwaki, Morr\'as, Mours, Mow-Lowry, Mozzon, Muciaccia, Mukherjee,
  Mukherjee, Mukherjee, Mukherjee, Mukherjee, Mukund, Mullavey, Munch, Mu\~niz,
  Murray, Musenich, Muusse, Nadji, Nagano, Nagano, Nagar, Nakamura, Nakano,
  Nakano, Nakashima, Nakayama, Napolano, Nardecchia, Narikawa, Naticchioni,
  Nayak, Nayak, Negishi, Neil, Neilson, Nelemans, Nelson, Nery, Neubauer,
  Neunzert, Ng, Ng, Nguyen, Nguyen, Nguyen, Quynh, Ni, Nichols, Nishizawa,
  Nissanke, Nitoglia, Nocera, Norman, North, Nozaki, Siles, Nuttall, Oberling,
  O'Brien, Obuchi, O'Dell, Oelker, Ogaki, Oganesyan, Oh, Oh, Oh, Ohashi,
  Ohishi, Ohkawa, Ohme, Ohta, Okada, Okutani, Okutomi, Olivetto, Oohara, Ooi,
  Oram, O'Reilly, Ormiston, Ormsby, Ortega, O'Shaughnessy, O'Shea, Oshino,
  Ossokine, Osthelder, Otabe, Ottaway, Overmier, Pace, Pagano, Page,
  Pagliaroli, Pai, Pai, Palamos, Palashov, Palomba, Pan, Pan, Panda, Pang,
  Pang, Pankow, Pannarale, Pant, Panther, Paoletti, Paoli, Paolone, Parisi,
  Park, Park, Parker, Pascucci, Pasqualetti, Passaquieti, Passuello, Patel,
  Pathak, Patricelli, Patron, Paul, Payne, Pedraza, Pegoraro, Pele, Arellano,
  Penn, Perego, Pereira, Pereira, Perez, P\'erigois, Perkins, Perreca,
  Perri\`es, Petermann, Petterson, Pfeiffer, Pham, Phukon, Piccinni, Pichot,
  Piendibene, Piergiovanni, Pierini, Pierro, Pillant, Pillas, Pilo, Pinard,
  Pinto, Pinto, Piotrzkowski, Piotrzkowski, Pirello, Pitkin, Placidi, Planas,
  Plastino, Pluchar, Poggiani, Polini, Pong, Ponrathnam, Popolizio, Porter,
  Poulton, Powell, Pracchia, Pradier, Prajapati, Prasai, Prasanna, Pratten,
  Principe, Prodi, Prokhorov, Prosposito, Prudenzi, Puecher, Punturo, Puosi,
  Puppo, P\"urrer, Qi, Quetschke, Quitzow-James, Raab, Raaijmakers, Radkins,
  Radulesco, Raffai, Rail, Raja, Rajan, Ramirez, Ramirez, Ramos-Buades, Rana,
  Rapagnani, Rapol, Ray, Raymond, Raza, Razzano, Read, Rees, Regimbau, Rei,
  Reid, Reid, Reitze, Relton, Renzini, Rettegno, Reza, Rezac, Ricci, Richards,
  Richardson, Richardson, Riemenschneider, Riles, Rinaldi, Rink, Rizzo,
  Robertson, Robie, Robinet, Rocchi, Rodriguez, Rolland, Rollins, Romanelli,
  Romano, Romel, Romero-Rodr\'{\i}guez, Romero-Shaw, Romie, Ronchini, Rosa,
  Rose, Rosi\ifmmode~\acute{n}\else \'{n}\fi{}ska, Ross, Rowan, Rowlinson, Roy,
  Roy, Roy, Rozza, Ruggi, Ryan, Sachdev, Sadecki, Sadiq, Sago, Saito, Saito,
  Sakai, Sakai, Sakellariadou, Sakuno, Salafia, Salconi, Saleem, Salemi,
  Samajdar, Sanchez, Sanchez, Sanchez, Sanchis-Gual, Sanders, Sanuy, Saravanan,
  Sarin, Sassolas, Satari, Sathyaprakash, Sato, Sato, Sauter, Savage, Sawada,
  Sawant, Sawant, Sayah, Schaetzl, Scheel, Scheuer, Schiworski, Schmidt,
  Schmidt, Schnabel, Schneewind, Schofield, Sch\"onbeck, Schulte, Schutz,
  Schwartz, Scott, Scott, Seglar-Arroyo, Sekiguchi, Sekiguchi, Sellers,
  Sengupta, Sentenac, Seo, Sequino, Sergeev, Setyawati, Shaffer, Shahriar,
  Shams, Shao, Sharma, Sharma, Shawhan, Shcheblanov, Shibagaki, Shikauchi,
  Shimizu, Shimoda, Shimode, Shinkai, Shishido, Shoda, Shoemaker, Shoemaker,
  ShyamSundar, Sieniawska, Sigg, Singer, Singh, Singh, Singha, Sintes, Sipala,
  Skliris, Slagmolen, Slaven-Blair, Smetana, Smith, Smith, Soldateschi, Somala,
  Somiya, Son, Soni, Soni, Sordini, Sorrentino, Sorrentino, Sotani, Soulard,
  Souradeep, Sowell, Spagnuolo, Spencer, Spera, Srinivasan, Srivastava,
  Srivastava, Staats, Stachie, Steer, Steinhoff, Steinlechner, Steinlechner,
  Stevenson, Stops, Stover, Strain, Strang, Stratta, Strunk, Sturani, Stuver,
  Sudhagar, Sudhir, Sugimoto, Suh, Sullivan, Summerscales, Sun, Sun, Sunil,
  Sur, Suresh, Sutton, Suzuki, Suzuki, Swinkels, Szczepa\ifmmode~\acute{n}\else
  \'{n}\fi{}czyk, Szewczyk, Tacca, Tagoshi, Tait, Takahashi, Takahashi,
  Takamori, Takano, Takeda, Takeda, Talbot, Talbot, Tanaka, Tanaka, Tanaka,
  Tanaka, Tanaka, Tanasijczuk, Tanioka, Tanner, Tao, Tao, Mart\'{\i}n, Taranto,
  Tasson, Telada, Tenorio, Terhune, Terkowski, Thirugnanasambandam, Thomas,
  Thomas, Thomas, Thompson, Thondapu, Thorne, Thrane, Tiwari, Tiwari, Tiwari,
  Toivonen, Toland, Tolley, Tomaru, Tomigami, Tomura, Tonelli, Torres-Forn\'e,
  Torrie, e~Melo, T\"oyr\"a, Trapananti, Travasso, Traylor, Trevor, Tringali,
  Tripathee, Troiano, Trovato, Trozzo, Trudeau, Tsai, Tsai, Tsang, Tsang, Tsao,
  Tse, Tso, Tsubono, Tsuchida, Tsukada, Tsuna, Tsutsui, Tsuzuki, Turbang,
  Turconi, Tuyenbayev, Ubhi, Uchikata, Uchiyama, Udall, Ueda, Uehara, Ueno,
  Ueshima, Unnikrishnan, Uraguchi, Urban, Ushiba, Utina, Vahlbruch, Vajente,
  Vajpeyi, Valdes, Valentini, Valsan, van Bakel, van Beuzekom, van~den Brand,
  Van Den~Broeck, Vander-Hyde, van~der Schaaf, van Heijningen, Vanosky, van
  Putten, van Remortel, Vardaro, Vargas, Varma, Vas\'uth, Vecchio, Vedovato,
  Veitch, Veitch, Venneberg, Venugopalan, Verkindt, Verma, Verma, Veske,
  Vetrano, Vicer\'e, Vidyant, Viets, Vijaykumar, Villa-Ortega, Vinet, Virtuoso,
  Vitale, Vo, Vocca, von Reis, von Wrangel, Vorvick, Vyatchanin, Wade, Wade,
  Wagner, Walet, Walker, Wallace, Wallace, Walsh, Wang, Wang, Wang, Ward,
  Warner, Was, Washimi, Washington, Watchi, Weaver, Webster, Weinert,
  Weinstein, Weiss, Weller, Wellmann, Wen, We\ss{}els, Wette, Whelan, White,
  Whiting, Whittle, Wilken, Williams, Williams, Williamson, Willis, Willke,
  Wilson, Winkler, Wipf, Wlodarczyk, Woan, Woehler, Wofford, Wong, Wu, Wu, Wu,
  Wu, Wysocki, Xiao, Xu, Yamada, Yamamoto, Yamamoto, Yamamoto, Yamamoto,
  Yamashita, Yamazaki, Yang, Yang, Yang, Yang, Yang, Yap, Yeeles, Yelikar,
  Ying, Yokogawa, Yokoyama, Yokozawa, Yoo, Yoshioka, Yu, Yu, Yuzurihara,
  Zadro\ifmmode~\dot{z}\else \.{z}\fi{}ny, Zanolin, Zeidler, Zelenova, Zendri,
  Zevin, Zhan, Zhang, Zhang, Zhang, Zhang, Zhang, Zhao, Zhao, Zhao, Zhao,
  Zheng, Zhou, Zhou, Zhu, Zhu, Zimmerman, Zlochower, Zucker, \&
  Zweizig}]{Abbot2023b}
Abbott, R., Abbott, T.~D., {et~al.} 2023,
  \href{http://dx.doi.org/10.1103/PhysRevX.13.011048}{\color{magenta}Phys. Rev.
  X}, 13, 011048

\bibitem[{Alexander \& Hopman(2009)}]{Tal_2009}
Alexander, T. \& Hopman, C. 2009,
  \href{http://dx.doi.org/10.1088/0004-637x/697/2/1861}{\color{magenta}The
  Astrophysical Journal}, 697, 1861–1869

\bibitem[{{Amaro-Seoane}(2018)}]{Pau2018LRR}
{Amaro-Seoane}, P. 2018,
  \href{http://dx.doi.org/10.1007/s41114-018-0013-8}{\color{magenta}Living
  Reviews in Relativity},
  \href{https://ui.adsabs.harvard.edu/abs/2018LRR....21....4A}{21, 4}

\bibitem[{Amaro-Seoane {et~al.}(2025)Amaro-Seoane, Arnau, \& Fullana~i
  Alfonso}]{AmaroSeoane2025b}
Amaro-Seoane, P., Arnau, J.~V., \& Fullana~i Alfonso, M.~J. 2025, arXiv
  preprint [\eprint{250420147A}]

\bibitem[{Amaro-Seoane {et~al.}(2017)Amaro-Seoane, Audley, Babak, Baker,
  Barausse, Bender, Berti, Binetruy, Born, Bortoluzzi, Camp, Caprini, Cardoso,
  Colpi, Conklin, Cornish, Cutler, Danzmann, Dolesi, Ferraioli, Ferroni,
  Fitzsimons, Gair, Bote, Giardini, Gibert, Grimani, Halloin, Heinzel, Hertog,
  Hewitson, Holley-Bockelmann, Hollington, Hueller, Inchauspe, Jetzer,
  Karnesis, Killow, Klein, Klipstein, Korsakova, Larson, Livas, Lloro, Man,
  Mance, Martino, Mateos, McKenzie, McWilliams, Miller, Mueller, Nardini,
  Nelemans, Nofrarias, Petiteau, Pivato, Plagnol, Porter, Reiche, Robertson,
  Robertson, Rossi, Russano, Schutz, Sesana, Shoemaker, Slutsky, Sopuerta,
  Sumner, Tamanini, Thorpe, Troebs, Vallisneri, Vecchio, Vetrugno, Vitale,
  Volonteri, Wanner, Ward, Wass, Weber, Ziemer, \& Zweifel}]{LISA_2017}
Amaro-Seoane, P., Audley, H., {et~al.} 2017, Laser Interferometer Space Antenna
  (ESA PUBLICATIONS DIVISION C/O ESTEC), submitted to ESA on January 13th in
  response to the call for missions for the L3 slot in the Cosmic Vision
  Programme

\bibitem[{Amaro~Seoane {et~al.}(2024)Amaro~Seoane, Lin, \&
  Tzanavaris}]{Pau_2024}
Amaro~Seoane, P., Lin, Y., \& Tzanavaris, K. 2024,
  \href{http://dx.doi.org/10.1103/PhysRevD.110.064011}{\color{magenta}Phys.
  Rev. D}, 110, 064011

\bibitem[{Amaro-Seoane \& Zhao(2025)}]{AmaroSeoane2025a}
Amaro-Seoane, P. \& Zhao, S.-D. 2025, arXiv preprint [\eprint{250410594A}]

\bibitem[{{Astropy Collaboration} {et~al.}(2022){Astropy Collaboration},
  {Price-Whelan}, {Lim}, {Earl}, {Starkman}, {Bradley}, {Shupe}, {Patil},
  {Corrales}, {Brasseur}, {N{\"o}the}, {Donath}, {Tollerud}, {Morris},
  {Ginsburg}, {Vaher}, {Weaver}, {Tocknell}, {Jamieson}, {van Kerkwijk},
  {Robitaille}, {Merry}, {Bachetti}, {G{\"u}nther}, {Aldcroft},
  {Alvarado-Montes}, {Archibald}, {B{\'o}di}, {Bapat}, {Barentsen},
  {Baz{\'a}n}, {Biswas}, {Boquien}, {Burke}, {Cara}, {Cara}, {Conroy},
  {Conseil}, {Craig}, {Cross}, {Cruz}, {D'Eugenio}, {Dencheva}, {Devillepoix},
  {Dietrich}, {Eigenbrot}, {Erben}, {Ferreira}, {Foreman-Mackey}, {Fox},
  {Freij}, {Garg}, {Geda}, {Glattly}, {Gondhalekar}, {Gordon}, {Grant},
  {Greenfield}, {Groener}, {Guest}, {Gurovich}, {Handberg}, {Hart},
  {Hatfield-Dodds}, {Homeier}, {Hosseinzadeh}, {Jenness}, {Jones}, {Joseph},
  {Kalmbach}, {Karamehmetoglu}, {Ka{\l}uszy{\'n}ski}, {Kelley}, {Kern},
  {Kerzendorf}, {Koch}, {Kulumani}, {Lee}, {Ly}, {Ma}, {MacBride}, {Maljaars},
  {Muna}, {Murphy}, {Norman}, {O'Steen}, {Oman}, {Pacifici}, {Pascual},
  {Pascual-Granado}, {Patil}, {Perren}, {Pickering}, {Rastogi}, {Roulston},
  {Ryan}, {Rykoff}, {Sabater}, {Sakurikar}, {Salgado}, {Sanghi}, {Saunders},
  {Savchenko}, {Schwardt}, {Seifert-Eckert}, {Shih}, {Jain}, {Shukla}, {Sick},
  {Simpson}, {Singanamalla}, {Singer}, {Singhal}, {Sinha}, {Sip{\H{o}}cz},
  {Spitler}, {Stansby}, {Streicher}, {{\v{S}}umak}, {Swinbank}, {Taranu},
  {Tewary}, {Tremblay}, {de Val-Borro}, {Van Kooten}, {Vasovi{\'c}}, {Verma},
  {de Miranda Cardoso}, {Williams}, {Wilson}, {Winkel}, {Wood-Vasey}, {Xue},
  {Yoachim}, {Zhang}, {Zonca}, \& {Astropy Project
  Contributors}}]{astropy+2022}
{Astropy Collaboration}, {Price-Whelan}, A.~M., {et~al.} 2022,
  \href{http://dx.doi.org/10.3847/1538-4357/ac7c74}{\color{magenta}\apj},
  \href{https://ui.adsabs.harvard.edu/abs/2022ApJ...935..167A}{935, 167}

\bibitem[{{Astropy Collaboration} {et~al.}(2018){Astropy Collaboration},
  {Price-Whelan}, {Sip{\H{o}}cz}, {G{\"u}nther}, {Lim}, {Crawford}, {Conseil},
  {Shupe}, {Craig}, {Dencheva}, {Ginsburg}, {VanderPlas}, {Bradley},
  {P{\'e}rez-Su{\'a}rez}, {de Val-Borro}, {Aldcroft}, {Cruz}, {Robitaille},
  {Tollerud}, {Ardelean}, {Babej}, {Bach}, {Bachetti}, {Bakanov}, {Bamford},
  {Barentsen}, {Barmby}, {Baumbach}, {Berry}, {Biscani}, {Boquien}, {Bostroem},
  {Bouma}, {Brammer}, {Bray}, {Breytenbach}, {Buddelmeijer}, {Burke},
  {Calderone}, {Cano Rodr{\'\i}guez}, {Cara}, {Cardoso}, {Cheedella}, {Copin},
  {Corrales}, {Crichton}, {D'Avella}, {Deil}, {Depagne}, {Dietrich}, {Donath},
  {Droettboom}, {Earl}, {Erben}, {Fabbro}, {Ferreira}, {Finethy}, {Fox},
  {Garrison}, {Gibbons}, {Goldstein}, {Gommers}, {Greco}, {Greenfield},
  {Groener}, {Grollier}, {Hagen}, {Hirst}, {Homeier}, {Horton}, {Hosseinzadeh},
  {Hu}, {Hunkeler}, {Ivezi{\'c}}, {Jain}, {Jenness}, {Kanarek}, {Kendrew},
  {Kern}, {Kerzendorf}, {Khvalko}, {King}, {Kirkby}, {Kulkarni}, {Kumar},
  {Lee}, {Lenz}, {Littlefair}, {Ma}, {Macleod}, {Mastropietro}, {McCully},
  {Montagnac}, {Morris}, {Mueller}, {Mumford}, {Muna}, {Murphy}, {Nelson},
  {Nguyen}, {Ninan}, {N{\"o}the}, {Ogaz}, {Oh}, {Parejko}, {Parley}, {Pascual},
  {Patil}, {Patil}, {Plunkett}, {Prochaska}, {Rastogi}, {Reddy Janga},
  {Sabater}, {Sakurikar}, {Seifert}, {Sherbert}, {Sherwood-Taylor}, {Shih},
  {Sick}, {Silbiger}, {Singanamalla}, {Singer}, {Sladen}, {Sooley},
  {Sornarajah}, {Streicher}, {Teuben}, {Thomas}, {Tremblay}, {Turner},
  {Terr{\'o}n}, {van Kerkwijk}, {de la Vega}, {Watkins}, {Weaver}, {Whitmore},
  {Woillez}, {Zabalza}, \& {Astropy Contributors}}]{astropy+2018}
{Astropy Collaboration}, {Price-Whelan}, A.~M., {et~al.} 2018,
  \href{http://dx.doi.org/10.3847/1538-3881/aabc4f}{\color{magenta}\aj},
  \href{https://ui.adsabs.harvard.edu/abs/2018AJ....156..123A}{156, 123}

\bibitem[{{Astropy Collaboration} {et~al.}(2013){Astropy Collaboration},
  {Robitaille}, {Tollerud}, {Greenfield}, {Droettboom}, {Bray}, {Aldcroft},
  {Davis}, {Ginsburg}, {Price-Whelan}, {Kerzendorf}, {Conley}, {Crighton},
  {Barbary}, {Muna}, {Ferguson}, {Grollier}, {Parikh}, {Nair}, {Unther},
  {Deil}, {Woillez}, {Conseil}, {Kramer}, {Turner}, {Singer}, {Fox}, {Weaver},
  {Zabalza}, {Edwards}, {Azalee Bostroem}, {Burke}, {Casey}, {Crawford},
  {Dencheva}, {Ely}, {Jenness}, {Labrie}, {Lim}, {Pierfederici}, {Pontzen},
  {Ptak}, {Refsdal}, {Servillat}, \& {Streicher}}]{astropy+2013}
{Astropy Collaboration}, {Robitaille}, T.~P., {et~al.} 2013,
  \href{http://dx.doi.org/10.1051/0004-6361/201322068}{\color{magenta}\aap},
  \href{https://ui.adsabs.harvard.edu/abs/2013A&A...558A..33A}{558, A33}

\bibitem[{{Bahcall} \& {Wolf}(1976)}]{BW_1976_1}
{Bahcall}, J.~N. \& {Wolf}, R.~A. 1976,
  \href{http://dx.doi.org/10.1086/154711}{\color{magenta}\apj},
  \href{https://ui.adsabs.harvard.edu/abs/1976ApJ...209..214B}{209, 214}

\bibitem[{{Bahcall} \& {Wolf}(1977)}]{BW_1977_2}
{Bahcall}, J.~N. \& {Wolf}, R.~A. 1977,
  \href{http://dx.doi.org/10.1086/155534}{\color{magenta}\apj},
  \href{https://ui.adsabs.harvard.edu/abs/1977ApJ...216..883B}{216, 883}

\bibitem[{Capuzzo-Dolcetta \& Sadun-Bordoni(2023)}]{Cap_Sad_2023}
Capuzzo-Dolcetta, R. \& Sadun-Bordoni, M. 2023,
  \href{http://dx.doi.org/10.1093/mnras/stad1317}{\color{magenta}Monthly
  Notices of the Royal Astronomical Society}, 522, 5828

\bibitem[{{Chassonnery} \& {Capuzzo-Dolcetta}(2021)}]{chacd21}
{Chassonnery}, P. \& {Capuzzo-Dolcetta}, R. 2021,
  \href{http://dx.doi.org/10.1093/mnras/stab1016}{\color{magenta}\mnras},
  \href{https://ui.adsabs.harvard.edu/abs/2021MNRAS.504.3909C}{504, 3909}

\bibitem[{{Chassonnery} {et~al.}(2019){Chassonnery}, {Capuzzo-Dolcetta}, \&
  {Mikkola}}]{chacd19}
{Chassonnery}, P., {Capuzzo-Dolcetta}, R., \& {Mikkola}, S. 2019,
  \href{https://ui.adsabs.harvard.edu/abs/2019arXiv191005202C}{\href{http://dx.doi.org/10.48550/arXiv.1910.05202}{\color{magenta}arXiv
  e-prints}, arXiv:1910.05202}

\bibitem[{{Davies} {et~al.}(2018){Davies}, {Alves}, {Cl{\'e}net}, {Lang-Bardl},
  {Nicklas}, {Pott}, {Ragazzoni}, {Tolstoy}, {Amico}, {Anwand-Heerwart},
  {Barboza}, {Barl}, {Baudoz}, {Bender}, {Bezawada}, {Bizenberger}, {Boland},
  {Bonifacio}, {Borgo}, {Buey}, {Chapron}, {Chemla}, {Cohen}, {Czoske},
  {D{\'e}o}, {Disseau}, {Dreizler}, {Dupuis}, {Fabricius}, {Falomo}, {Fedou},
  {F{\"o}rster Schreiber}, {Garrel}, {Geis}, {Gemperlein}, {Gendron}, {Genzel},
  {Gillessen}, {Gl{\"u}ck}, {Grupp}, {Hartl}, {H{\"a}user}, {Hess},
  {Hofferbert}, {Hopp}, {H{\"o}rmann}, {Hubert}, {Huby}, {Huet}, {Hutterer},
  {Ives}, {Janssen}, {Jellema}, {Kausch}, {Kerber}, {Kravcar}, {Le Ruyet},
  {Leschinski}, {Mandla}, {Manhart}, {Massari}, {Mei}, {Merlin}, {Mohr},
  {Monna}, {Muench}, {M{\"u}ller}, {Musters}, {Navarro}, {Neumann}, {Neumayer},
  {Niebsch}, {Plattner}, {Przybilla}, {Rabien}, {Ramlau}, {Ramos}, {Ramsay},
  {Rhode}, {Richter}, {Richter}, {Rix}, {Rodeghiero}, {Rohloff},
  {Rosensteiner}, {Rousset}, {Schlichter}, {Schubert}, {Sevin}, {Stuik},
  {Sturm}, {Thomas}, {Tromp}, {Verdoes-Kleijn}, {Vidal}, {Wagner}, {Wegner},
  {Zeilinger}, {Ziegleder}, {Ziegler}, \& {Zins}}]{MICADO_ELT}
{Davies}, R., {Alves}, J., {et~al.} 2018, in Society of Photo-Optical
  Instrumentation Engineers (SPIE) Conference Series, Vol. 10702, Ground-based
  and Airborne Instrumentation for Astronomy VII, ed. C.~J. {Evans},
  L.~{Simard}, \& H.~{Takami},
  \href{https://ui.adsabs.harvard.edu/abs/2018SPIE10702E..1SD}{107021S}

\bibitem[{{Frank} \& {Rees}(1976)}]{frank_rees_1976}
{Frank}, J. \& {Rees}, M.~J. 1976,
  \href{http://dx.doi.org/10.1093/mnras/176.3.633}{\color{magenta}\mnras},
  \href{https://ui.adsabs.harvard.edu/abs/1976MNRAS.176..633F}{176, 633}

\bibitem[{Freitag {et~al.}(2006)Freitag, Amaro‐Seoane, \&
  Kalogera}]{Freitag_2006}
Freitag, M., Amaro‐Seoane, P., \& Kalogera, V. 2006,
  \href{http://dx.doi.org/10.1086/506193}{\color{magenta}The Astrophysical
  Journal}, 649, 91–117

\bibitem[{{Ghez} {et~al.}(2003){Ghez}, {Duch{\^e}ne}, {Matthews}, {Hornstein},
  {Tanner}, {Larkin}, {Morris}, {Becklin}, {Salim}, {Kremenek}, {Thompson},
  {Soifer}, {Neugebauer}, \& {McLean}}]{Ghez_2003}
{Ghez}, A.~M., {Duch{\^e}ne}, G., {et~al.} 2003,
  \href{http://dx.doi.org/10.1086/374804}{\color{magenta}\apjl},
  \href{https://ui.adsabs.harvard.edu/abs/2003ApJ...586L.127G}{586, L127}

\bibitem[{{Ghez} {et~al.}(2008){Ghez}, {Salim}, {Weinberg}, {Lu}, {Do}, {Dunn},
  {Matthews}, {Morris}, {Yelda}, {Becklin}, {Kremenek}, {Milosavljevic}, \&
  {Naiman}}]{Ghez_2008}
{Ghez}, A.~M., {Salim}, S., {et~al.} 2008,
  \href{http://dx.doi.org/10.1086/592738}{\color{magenta}\apj},
  \href{https://ui.adsabs.harvard.edu/abs/2008ApJ...689.1044G}{689, 1044}

\bibitem[{Gillessen {et~al.}(2017)Gillessen, Plewa, Eisenhauer, Sari, Waisberg,
  Habibi, Pfuhl, George, Dexter, von Fellenberg, Ott, \&
  Genzel}]{Gillessen_2017}
Gillessen, S., Plewa, P.~M., {et~al.} 2017,
  \href{http://dx.doi.org/10.3847/1538-4357/aa5c41}{\color{magenta}\apj}, 837,
  30

\bibitem[{{GRAVITY Collaboration}(2017)}]{GRAVITY_2017}
{GRAVITY Collaboration}. 2017,
  \href{http://dx.doi.org/10.1051/0004-6361/201730838}{\color{magenta}\aap},
  602, A94

\bibitem[{{GRAVITY Collaboration}(2018a)}]{GRAVITY_2018}
{GRAVITY Collaboration}. 2018a,
  \href{http://dx.doi.org/10.1051/0004-6361/201833718}{\color{magenta}\aap},
  615, L15

\bibitem[{{GRAVITY Collaboration}(2019)}]{GRAVITY_2019}
{GRAVITY Collaboration}. 2019,
  \href{http://dx.doi.org/10.1051/0004-6361/201935656}{\color{magenta}\aap},
  625, L10

\bibitem[{{GRAVITY Collaboration}(2020)}]{GRAVITY_2020}
{GRAVITY Collaboration}. 2020,
  \href{http://dx.doi.org/10.1051/0004-6361/202037813}{\color{magenta}\aap},
  636, L5

\bibitem[{{GRAVITY+ Collaboration}(2022)}]{GRAVITY+}
{GRAVITY+ Collaboration}. 2022,
  \href{http://dx.doi.org/10.18727/0722-6691/5285}{\color{magenta}Published in
  The Messenger vol. 189}, pp. 17-22, December 2022.

\bibitem[{{GRAVITY Collaboration}(2022)}]{GRAVITY_2022}
{GRAVITY Collaboration}. 2022,
  \href{http://dx.doi.org/10.1051/0004-6361/202142465}{\color{magenta}\aap},
  657, L12

\bibitem[{{GRAVITY Collaboration}(2023)}]{GRAVITY_2023}
{GRAVITY Collaboration}. 2023,
  \href{http://dx.doi.org/10.1051/0004-6361/202245132}{\color{magenta}A\&A},
  672, A63

\bibitem[{{GRAVITY Collaboration}(2024)}]{GRAVITY_2024}
{GRAVITY Collaboration}. 2024,
  \href{http://dx.doi.org/10.1051/0004-6361/202452274}{\color{magenta}A\&A},
  692, A242

\bibitem[{Harris {et~al.}(2020)Harris, Millman, van~der Walt, Gommers,
  Virtanen, Cournapeau, Wieser, Taylor, Berg, Smith, Kern, Picus, Hoyer, van
  Kerkwijk, Brett, Haldane, del R{\'i}o, Wiebe, Peterson, G{\'e}rard-Marchant,
  Sheppard, Reddy, Weckesser, Abbasi, Gohlke, \& Oliphant}]{harris+2020Numpy}
Harris, C.~R., Millman, K.~J., {et~al.} 2020,
  \href{http://dx.doi.org/10.1038/s41586-020-2649-2}{\color{magenta}Nature},
  585, 357

\bibitem[{Hintze \& Nelson(1998)}]{Hintze_1998}
Hintze, J.~L. \& Nelson, R.~D. 1998,
  \href{http://dx.doi.org/10.1080/00031305.1998.10480559}{\color{magenta}The
  American Statistician}, 52, 181

\bibitem[{{Hunter}(2007)}]{hunter2007}
{Hunter}, J.~D. 2007,
  \href{http://dx.doi.org/10.1109/MCSE.2007.55}{\color{magenta}Computing in
  Science and Engineering},
  \href{https://ui.adsabs.harvard.edu/abs/2007CSE.....9...90H}{9, 90}

\bibitem[{{Kluyver} {et~al.}(2016){Kluyver}, {Ragan-Kelley}, {P{\'e}rez},
  {Granger}, {Bussonnier}, {Frederic}, {Kelley}, {Hamrick}, {Grout}, {Corlay},
  {Ivanov}, {Avila}, {Abdalla}, {Willing}, \& {Jupyter Development
  Team}}]{kluyver+2016}
{Kluyver}, T., {Ragan-Kelley}, B., {et~al.} 2016, in IOS Press, 87--90

\bibitem[{Merritt {et~al.}(2010)Merritt, Alexander, Mikkola, \&
  Will}]{Merritt_2010}
Merritt, D., Alexander, T., {et~al.} 2010,
  \href{http://dx.doi.org/10.1103/physrevd.81.062002}{\color{magenta}Physical
  Review D}, 81

\bibitem[{Merritt {et~al.}(2007)Merritt, Berczik, \& Laun}]{Merritt_2007}
Merritt, D., Berczik, P., \& Laun, F. 2007,
  \href{http://dx.doi.org/10.1086/510294}{\color{magenta}The Astronomical
  Journal}, 133, 553

\bibitem[{{Mikkola} \& {Tanikawa}(1999{\natexlab{a}})}]{miktan99a}
{Mikkola}, S. \& {Tanikawa}, K. 1999{\natexlab{a}},
  \href{http://dx.doi.org/10.1046/j.1365-8711.1999.02982.x}{\color{magenta}\mnras},
  \href{https://ui.adsabs.harvard.edu/abs/1999MNRAS.310..745M}{310, 745}

\bibitem[{{Mikkola} \& {Tanikawa}(1999{\natexlab{b}})}]{miktan99b}
{Mikkola}, S. \& {Tanikawa}, K. 1999{\natexlab{b}},
  \href{http://dx.doi.org/10.1023/A:1008368322547}{\color{magenta}Celestial
  Mechanics and Dynamical Astronomy},
  \href{https://ui.adsabs.harvard.edu/abs/1999CeMDA..74..287M}{74, 287}

\bibitem[{Mora \& Will(2004)}]{mowill_2004}
Mora, T. \& Will, C.~M. 2004,
  \href{http://dx.doi.org/10.1103/PhysRevD.69.104021}{\color{magenta}Phys. Rev.
  D}, 69, 104021

\bibitem[{{Peebles}(1972)}]{Peebles_1972}
{Peebles}, P.~J.~E. 1972,
  \href{http://dx.doi.org/10.1086/151797}{\color{magenta}\apj},
  \href{https://ui.adsabs.harvard.edu/abs/1972ApJ...178..371P}{178, 371}

\bibitem[{{Preto} \& {Amaro-Seoane}(2010)}]{Preto_Pau_2010}
{Preto}, M. \& {Amaro-Seoane}, P. 2010,
  \href{http://dx.doi.org/10.1088/2041-8205/708/1/L42}{\color{magenta}\apjl},
  \href{https://ui.adsabs.harvard.edu/abs/2010ApJ...708L..42P}{708, L42}

\bibitem[{{Reid} \& {Brunthaler}(2020)}]{Reid_2020}
{Reid}, M.~J. \& {Brunthaler}, A. 2020,
  \href{http://dx.doi.org/10.3847/1538-4357/ab76cd}{\color{magenta}\apj},
  \href{https://ui.adsabs.harvard.edu/abs/2020ApJ...892...39R}{892, 39}

\bibitem[{{Rein} \& {Liu}(2012)}]{rein.liu2012}
{Rein}, H. \& {Liu}, S.~F. 2012,
  \href{http://dx.doi.org/10.1051/0004-6361/201118085}{\color{magenta}\aap},
  \href{https://ui.adsabs.harvard.edu/abs/2012A&A...537A.128R}{537, A128}

\bibitem[{{Rein} \& {Spiegel}(2015)}]{rein.spiegel2015}
{Rein}, H. \& {Spiegel}, D.~S. 2015,
  \href{http://dx.doi.org/10.1093/mnras/stu2164}{\color{magenta}\mnras},
  \href{https://ui.adsabs.harvard.edu/abs/2015MNRAS.446.1424R}{446, 1424}

\bibitem[{{Sabha} {et~al.}(2012){Sabha}, {Eckart, A.}, {Merritt, D.},
  {Zamaninasab, M.}, {Witzel, G.}, {García-Marín, M.}, {Jalali, B.}, {, M.
  Valencia-S.}, {Yazici, S.}, {Buchholz, R.}, {Shahzamanian, B.}, {Rauch, C.},
  {Horrobin, M.}, \& {Straubmeier, C.}}]{Sabha_2012}
{Sabha}, N., {Eckart, A.}, {et~al.} 2012,
  \href{http://dx.doi.org/10.1051/0004-6361/201219203}{\color{magenta}A\&A},
  545, A70

\bibitem[{{Sch{\"o}del} {et~al.}(2002){Sch{\"o}del}, {Ott}, {Genzel},
  {Hofmann}, {Lehnert}, {Eckart}, {Mouawad}, {Alexander}, {Reid}, {Lenzen},
  {Hartung}, {Lacombe}, {Rouan}, {Gendron}, {Rousset}, {Lagrange}, {Brandner},
  {Ageorges}, {Lidman}, {Moorwood}, {Spyromilio}, {Hubin}, \&
  {Menten}}]{Schodel_2002}
{Sch{\"o}del}, R., {Ott}, T., {et~al.} 2002,
  \href{http://dx.doi.org/10.1038/nature01121}{\color{magenta}\nat},
  \href{https://ui.adsabs.harvard.edu/abs/2002Natur.419..694S}{419, 694}

\bibitem[{{Tamayo} {et~al.}(2020){Tamayo}, {Rein}, {Shi}, \&
  {Hernandez}}]{tamayo+2020}
{Tamayo}, D., {Rein}, H., {et~al.} 2020,
  \href{http://dx.doi.org/10.1093/mnras/stz2870}{\color{magenta}\mnras},
  \href{https://ui.adsabs.harvard.edu/abs/2020MNRAS.491.2885T}{491, 2885}

\bibitem[{{Virtanen} {et~al.}(2020){Virtanen}, {Gommers}, {Oliphant},
  {Haberland}, {Reddy}, {Cournapeau}, {Burovski}, {Peterson}, {Weckesser},
  {Bright}, {van der Walt}, {Brett}, {Wilson}, {Millman}, {Mayorov}, {Nelson},
  {Jones}, {Kern}, {Larson}, {Carey}, {Polat}, {Feng}, {Moore}, {VanderPlas},
  {Laxalde}, {Perktold}, {Cimrman}, {Henriksen}, {Quintero}, {Harris},
  {Archibald}, {Ribeiro}, {Pedregosa}, {van Mulbregt}, \& {SciPy 1. 0
  Contributors}}]{virtanen+2020}
{Virtanen}, P., {Gommers}, R., {et~al.} 2020,
  \href{http://dx.doi.org/10.1038/s41592-019-0686-2}{\color{magenta}Nature
  Methods}, \href{https://ui.adsabs.harvard.edu/abs/2020NatMe..17..261V}{17,
  261}

\bibitem[{{Waisberg} {et~al.}(2018){Waisberg}, {Dexter}, {Gillessen}, {Pfuhl},
  {Eisenhauer}, {Plewa}, {Baub{\"o}ck}, {Jimenez-Rosales}, {Habibi}, {Ott},
  {von Fellenberg}, {Gao}, {Widmann}, \& {Genzel}}]{Waisb_2018}
{Waisberg}, I., {Dexter}, J., {et~al.} 2018,
  \href{http://dx.doi.org/10.1093/mnras/sty476}{\color{magenta}\mnras},
  \href{https://ui.adsabs.harvard.edu/abs/2018MNRAS.476.3600W}{476, 3600}

\bibitem[{{W}es {M}c{K}inney(2010)}]{pandas2010}
{{W}es {M}c{K}inney. 2010, in {P}roceedings of the 9th {P}ython in {S}cience
  {C}onference, ed. {S}t\'efan van~der {W}alt \& {J}arrod {M}illman, }{56 --
  61}

\bibitem[{Will(1993)}]{Will_1993}
Will, C.~M. 1993, Theory and Experiment in Gravitational Physics (Cambridge
  University Press)

\bibitem[{Will(2014)}]{Will_2014}
Will, C.~M. 2014,
  \href{http://dx.doi.org/10.12942/lrr-2014-4}{\color{magenta}Living Reviews in
  Relativity}, 17

\bibitem[{Will {et~al.}(2023)Will, Naoz, Hees, Tucker, Zhang, Do, \&
  Ghez}]{Will_2023}
Will, C.~M., Naoz, S., {et~al.} 2023,
  \href{http://dx.doi.org/10.3847/1538-4357/ad09b3}{\color{magenta}The
  Astrophysical Journal}, 959, 58

\bibitem[{Zhang \& Amaro-Seoane(2024)}]{Zhang_Pau_2023}
Zhang, F. \& Amaro-Seoane, P. 2024,
  \href{http://dx.doi.org/10.3847/1538-4357/ad0f1a}{\color{magenta}The
  Astrophysical Journal}, 961, 232

\bibitem[{{Zhang} \& {Iorio}(2017)}]{Zhang_Iorio_2017}
{Zhang}, F. \& {Iorio}, L. 2017,
  \href{http://dx.doi.org/10.3847/1538-4357/834/2/198}{\color{magenta}\apj},
  \href{https://ui.adsabs.harvard.edu/abs/2017ApJ...834..198Z}{834, 198}

\end{thebibliography}



\end{document}